\documentclass[aps,showpacs,showkeys,nofootinbib,twocolumn]{revtex4}
\usepackage{epsfig,amssymb,bm,graphics,color}

\newcommand*{\ee}{e^+e^-}

\newcommand*{\spt}{\sigma_{\perp}}
\newcommand*{\spl}{\sigma_{\parallel}}

\begin{document} {\normalsize }
 \title{
Non-linear Breit-Wheeler process with linearly polarized beams
}
%
 \author{A~I.~Titov}
 \affiliation{
 Bogoliubov Laboratory of Theoretical Physics, JINR, Dubna 141980, Russia,\\%
 Email:atitov@theor.jinr.ru}
\author{B.~K\"ampfer}
\affiliation{
Helmholtz-Zentrum  Dresden-Rossendorf, 01314 Dresden, Germany\\%
Institut f\"ur Theoretische Physik, TU~Dresden, 01062 Dresden, Germany}

 \begin{abstract}
We study the non-linear Breit-Wheeler process
$\vec \gamma' + \vec L \to e^+ + e^-$
in the interaction of linearly polarized probe photons ($\vec \gamma'$)
with a linearly polarized laser beam ($\vec L$).
In particular, we consider the asymmetry of the total cross section
and the azimuthal electron distributions when
the polarizations of the photon and laser beams in the initial state are mutually
perpendicular 
or parallel.  
Considering intense laser beams and the strong field asymptotic
we explore essentially the multi-photon dynamics.
The asymmetry exhibits some non-monotonic behavior depending on initial
kinematic conditions; it depends sensitively on the laser pulse duration.
Our results provide additional knowledge for studying non-linear
multi-photon effects in quantum electrodynamics
and may be used in planning experiments in upcoming laser facilities.
 \end{abstract}

\maketitle

\section{Introduction}

The study of elementary photon-electron interaction processes
in strong electromagnetic fields in the laboratory is enabled,
to a large extent, by
facilities which provide intense and ultra-intense laser beams.
Existing installations and forthcoming  high-power laser projects
allow testing quantum electrodynamics (QED)
as a pillar of the Standard Model in the non-linear
regime. Besides XFEL beams, the optical lasers play a key role.
Among the latter ones, the ELI-Beam Lines~\cite{C1} and
ELI-Nuclear Physics~\cite{C2} facilities are widely discussed now,
to be seen in the context of many other projects, cf.~\cite{Meuren:2020nbw}.
For a review on most known upcoming petawatt
and exawatt laser projects, see~\cite{C3} for instance.

When speaking on ``elementary QED interaction processes" we have
in mind Schwinger pair production, Breit-Wheeler and trident (triple)
pair production, and Compton scattering. While ranging from elusive to
not much probed to fairly well known, these fundamental phenomena
can be considered as corner stones of QED which deserve in-depth
investigations in their own right, in particular in the strong-field regime.
In the present work we focus on some details of polarization effects in the
non-linear Breit-Wheeler (BW) process.  Within the Furry picture,
the non-linear BW process
refers to the decay of a linearly polarized probe photon
$\vec \gamma'$ with four-momentum $k'(\omega',{\mathbf k}')$
into an electron-positron ($e^+ e^-$) pair while traversing a
linearly polarized laser pulse
(characterized by the central frequency $\omega$, wave four-vector $k$ and
polarization four-vector $\varepsilon$ with $\varepsilon \cdot k = 0$, where the dot stands
for the scalar product), symbolically $\vec \gamma' \to e_L^- + e_L^+$. Here,
the label ``$L$" points to the laser-dressed $e^\pm$ states.
Alternatively, one can characterize the reaction under consideration by
$\vec \gamma' + \vec L \to e^+ + e^-$.
Facilities of the probe photon beam polarization
for the multi-photon BW process are discussed
in \cite{Wan:2020zet,Li:2019oxr,Blackburn:2018ghi}, for instance.
Polarized multi-GeV photon beams are in operation worldwide,
cf.~\cite{Ambrozewicz:2019xly,Adhikari:2020cvz}.

Various laser polarizations can be accomplished customarily.
In the plane-wave approximation, one can further distinguish monochromatic
laser beams, formally with an infinitely long duration
(that is the infinite-pulse approximation [IPA]), or a pulse of finite duration
(that is the finite-pulse approximation [FPA]). In the latter case,
the bandwidth effects cause a distribution of frequencies around the  central
one, as evidenced by the power spectrum of the pulse. The IPA case
has been analyzed in depth by Reiss~\cite{Reissold} and
the Ritus group~\cite{RitusGroup} some time ago,
and summarized in a well-known review paper~\cite{Ritus-79}.
For completeness, we mention
the review papers~\cite{Mourou1,Piazza1,Narozhny2015v}, and also recent
publications~\cite{Mourou2,Piazza2,Titov2020,Heinzl2020}.

The weak-field BW pair production $\gamma' + \gamma \to e^+ + e^-$
is a threshold process requiring $s > 4 m^2$
(where $s$ and $m$ are the square of the total energy
in the center of mass system (c.m.s.) and the electron mass, respectively),
therefore, for its analysis it is natural to use two dimensionless
relativistic and invariant variables.
One is the reduced field intensity $\xi$ related to the intensity
of background field potential $a$, and electron charge $e$, $\xi=ea/m$ ~\cite{LL},
and the threshold variable $\zeta \equiv 4m^2/s$
\cite{TitovPEPAN} instead of the Mandelstam variable $s$.
The region $\zeta>1$ automatically
selects the sub-threshold multi-photon regime,
where the simultaneous participation of a multitude of
photons in the laser beam via $\gamma' + n \gamma \to e^+ + e^-$
enables the pair production.
Instead of $\zeta$, one can equally well use
the quantum-nonlinearity parameter $\kappa$
(known as a Ritus variable),
related to $\zeta$ by $\kappa=2\xi/\zeta$.

The  SLAC experiment E-144~\cite{E-144}
has tested the sub-threshold multi-photon regime
with $n > 3$ at $\xi \leq 0.35$
by the trident process $e^- + L \to e^- + e^+ + e^-$
which combines the sub-processes of non-linear Compton back-scattering
$e^- + L \to e^- + \gamma'$ and non-linear BW pair production.
The envisaged LUXE experiment \cite{LUXE,Abramowicz:2019gvx}
will probe the non-linear BW and trident processes
at $\xi > 1$ at the precision level.
For further prospects, see \cite{Meuren:2020nbw},
in particular w.r.t.\ FACET-II \cite{FACET-II}.

Given the present and future experimental research opportunities,
the theoretical basis must be developed in more detail.
In early works \cite{Reissold,RitusGroup,Ritus-79},
it was found that the probability of electron-positron BW
pair production depends on the mutual polarization of
the probe photon and the laser background field.
For example, different probabilities ($W$) (or cross sections ($\sigma$))
have been calculated for
the non-linear BW process for mutual polarizations being either
perpendicular $(\perp)$ or parallel ($\parallel$).
In the case of a monochromatic
background field and for asymptotically large values $\xi \gg 1$,
Ref.~\cite{Ritus-79} predicts a ratio of
$\spt/\spl$ equal to 2 and 3/2
for $\kappa\ll1$ and $\kappa\gg1$, respectively.
Some definite peculiarities in the differential distributions of
positrons depending on the mutually
linear polarizations of the laser pulse and probe photon beam at finite
values of $\xi$ were considered in \cite{Krajewska:2012eb}.
The non-linear BW pair production in short laser pulses was studied
in \cite{Meuren:2014uia} in a wide region of $\xi$ and $\kappa$
by employing a polarization-operator approach.
An exponential decrease of the probability of $\ee$ pair creation
with decreasing values of $\kappa$ was found in the asymptotic region of
$\xi\gg1,\,\kappa\ll1$.
The observed decrease is even stronger than predicted
for the constant-cross field, while maintaining
the same ratio $ W_\perp/W_\parallel$ as in~\cite{Ritus-79}.

Another example of $\ee$ BW pair production at relatively high field
strengths corresponding to $\xi=1\dots5$ and the energy of the probe photon
$\omega'=13$~GeV was analyzed in \cite{Wistisen:2020rsq}
within a semi-classical approach to BW $\ee$ pair production.
It was found that, at $\xi=2$ and $\kappa=0.3 \dots 0.4$,
the relative probabilities of $\ee$ pair production with
different photon polarizations are $W_{\perp}=2.04\dots2.05$.

The difference between $\spt$ and $\spl$ leads to a finite
asymmetry $ {\cal A} = (\spt- \spl)/(\spt + \spl)$,
which has not yet been subject of an independent research in itself.
For example, the asymptotic prediction of \cite{Ritus-79} at $\xi \gg 1$
leads to variation of the asymmetry in the interval
${\cal A}=1/3$ to 1/5 for $\kappa\ll1$ and $\kappa\gg1$, respectively.

In present work we analyze the asymmetry ${\cal A}$ in two regions:
(i) at medium-strong ($\xi \leq 1$) and
(ii) ultra-strong ($\xi \gg 1$)
laser fields, respectively.
In the case of $\xi<1$, the beam duration
(or the number of e.m. field oscillations in the pulse)
is important \cite{Titov2020,Heinzl2020,Meuren:2014uia}. Therefore,
we analyze the asymmetry as a function of
$\xi$ and $\zeta$ for different pulse legths, using our formulation
developed in \cite{Titov2020}. We show that, in this region, the asymmetry
may vary within the interval from ${\cal A}\simeq 0$ to ${\cal A}\simeq 1$
where the cross section can acquire values from $\spl\simeq\spt$ to
$\spl\ll\spt$.

In the case of a strong laser field characterized by $\xi \gg 1$, the dominant
contribution to the probability of $\ee$ pair production comes from
the central part of the pulse envelope \cite{TitovPRA}.
Therefore, if the number of e.m.\ field oscillations exceeds unity,
the pulse duration becomes insignificant
and one can use the constant-crossed field approximation
in a wide region of $\zeta\,(\kappa)$ which excludes
the appearance of new parameters and assumptions. In both cases,
we analyze the dependence of ${\cal A}$ on the threshold parameter $\zeta$
and on the azimuthal angle of the outgoing electron (positron).

Our paper is organized as follows. In Sect.~II, for completeness, we
recall the laser field model. The deployed basic formulations of the cases
$\xi \leq 1$ and $\xi \gg 1$ are presented in Sects.~III and V.
The respective numerical results are discussed in Sects.~ IV and VI.
Our summary is given in Sect.~VII

\section{The background field model}

 We suppose the external, linearly polarized background
 (laser pulse) field is
 determined by the electromagnetic (e.m.) four-potential
 in the axial gauge $A=(0,{\mathbf A})$
 as ${\mathbf E}=-\partial {\mathbf A}/\partial t$:
 \begin{eqnarray}
  \mathbf{A}(\phi) = f(\phi) \left[ \mathbf{a}\cos(\phi)\right]~.
  \label{I1}
 \end{eqnarray}
The quantity
$\phi=k\cdot x$ is the invariant phase with four-wave vector
$k=(\omega, \mathbf{k})$, obeying the null field property
$k^2 = k \cdot k=0$
implying $\omega = \vert\mathbf{k}\vert$,
 $ \mathbf{a} \equiv \mathbf{a}_{(x)}$;
 $|\mathbf{a}|^2=a^2$;
transversality means $\mathbf{k} \mathbf{a}=0$ in the present gauge.
For the sake of definiteness,
the envelope function $f(\phi)$ is chosen as hyperbolic secant:
 \begin{eqnarray}
 f(\phi)=\frac{1}{\cosh\frac{\phi}{\Delta}} .
 \label{I3}
 \end{eqnarray}
 The dimensionless quantity $\Delta$ is related to the pulse duration
 $2\Delta=2\pi N$, where $N$ has the meaning of the number of cycles in
 the laser pulse.
 It is related to the time duration of the pulse $\tau=2N/\omega$.
$N < 1$ means sub-cycle pulses
(for the dependence of some observables on the envelope shape, see,
for example \cite{TitovPEPAN}).

\begin{figure}[th]
\includegraphics[width=0.6\columnwidth]{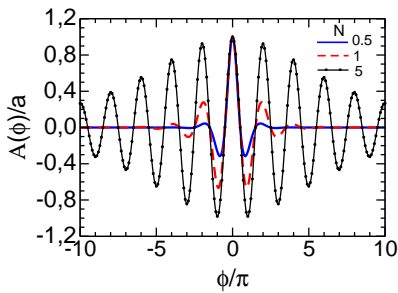} 
\caption{\small{(Color online)
The electromagnetic potential as a function of
invariant phase $\phi=k\cdot x$
for different values of the pulse duration, $N = 1/2$ (blue solid curve),
1 (dashed red curve) and 5 (black dash-dotted curve).
 \label{Fig:A1}}}
 \end{figure}

For an illustration, Fig.\ref{Fig:A1} exhibits
the e.m.\ potential $A$ of the pulse as a function of
invariant phase $\phi$ for different values of the parameter
$N=$0.5, 1 and 5 shown by solid, dashed and the dash-dotted curves,
respectively. These values of $N$ are used in further analysis.
The case of $N=0.5$ corresponds to the sub-cycle  pulse.

The cross section of $\ee$-pair production includes
a normalization factor $N_0$ which is related
to the average square of the e.m.\ strength
and is expressed through the envelope functions as
\begin{eqnarray}
N_0&=&\frac{1}{2 \pi}\int\limits_{-\infty}^{\infty}
d\phi \left(f^2(\phi)+{f'}^2(\phi)\right)\,\cos^2\phi
\label{I4}
\end{eqnarray}
with the asymptotic value
$N_0 \approx \Delta / 2 \pi$ at $\Delta / \pi \gg 1$.
We use natural units with
 $c=\hbar=1$, $e^2/4\pi = \alpha \approx 1/137.036$.

\section{Basics of cross section and asymmetry, \boldmath$\xi \leq 1$}

 As mentioned above,
 we consider essentially multi-photon events, where
 a finite number of laser photons is involved simultaneously in
 the $\ee$ pair production. This allows for sub-threshold
 $\ee$ pair production at $s < s_{\rm thr}$ or $\zeta > 1$.
 In this section we analyze the dependence of
 cross sections on $\zeta$ and on
 the e.m.\ field intensity which is described
 by the reduced field intensity parameter
 $\xi^2$ for different mutual polarizations of the incident photon and
 the laser beam.

 The differential cross sections read
\begin{eqnarray}
\frac{d \sigma_{i}}{d\phi_{e} }
=\frac{2\alpha^2}{m^2\xi\kappa N_0}
\,\int\limits_\zeta^{\infty}\,d\ell \,v(\ell) \,
\int\limits_{-1}^{1} d\cos\theta_{e}
\, w_i{(\ell)}~,
\label{II1}
\end{eqnarray}
where $i=\perp,\,\parallel$, $\ell$ is an auxiliary continuous variable,
$w_i(\ell)$ denotes the partial probability of the process process.
The azimuthal angle of the outgoing electron, $\phi_{e}$, is defined as
 $\cos\phi_{e}={\mathbf a_x}{\mathbf p}_{e}/a |{\mathbf p}_{e}|$.
 It is related to the azimuthal angle of the positron by
 $\phi_{e^+}=\phi_{e} + \pi$.
 Furthermore,  $\theta_{e}$ is the polar angle
 of the outgoing electron, $v$ is the electron (positron) velocity
 in the center of mass system (c.m.s.).

 The lower limit of the integral over the variable $\ell$
 is the threshold parameter $\zeta$.
 The region of $\zeta<1$
 corresponds to the above-threshold $\ee$ pair production, while
 the region of $\zeta>1$ is for the sub-threshold pair production
 enabled by multi-photon and bandwidth effects.
 We keep our notation in \cite{Titov2020} and
 denote by $k(\omega,{\mathbf k})$,
 $k'(\omega',{\mathbf k}')$,
 $p(E,{\mathbf p})$ and $p'(E',{\mathbf p}')$ the four-momenta of the
 background (laser) field, the
 incoming probe photon, the outgoing positron and the outgoing electron,
 respectively.  The important variables $s$, $v$ and $u$
 are determined by $s={2k\cdot k' }= 2(\omega'\omega -{\mathbf k}'{\mathbf k})$
 (with $\mathbf k' \mathbf k = - \omega' \omega$ for head-on geometry),
 $v^2=(\ell s-4m^2)/\ell s$,
 $u\equiv(k'\cdot k)^2/\left(4(k\cdot p)(k\cdot p')\right)
 =1/(1-v^2\cos^2\theta_e)$.
 The Ritus variable $\kappa=\xi(k\cdot k')/m^2$, is related to $\zeta$ by
 $\kappa=2\xi/\zeta$.
 Note the identity
 \begin{eqnarray}
\,v(\ell) \,
\int\limits_{-1}^{1} d\cos\theta_{e}=
\int\limits_{1}^{u_{\ell}} \frac{du}{u\sqrt{u(u-1)}}
\label{II2}
\end{eqnarray}
 with $u_{\ell}=\ell/\zeta$. The normalization factor $N_0$ is given by
 Eq.~(\ref{I4}).

 In cases, where the incident probe photon polarization plane is parallel ($\parallel$)
 or perpendicular ($\perp$) to the laser beam polarization, the partial
 probabilities $w_i(\ell)$ have the following form:
\begin{eqnarray}
&& w_{\parallel}(\ell) =
 \xi^2(u-1)
 \left(|\widetilde A_1(\ell)|^2 -
 {\rm Re}[\widetilde A_0(\ell)\widetilde A_2^*(\ell)\right)
 \nonumber\\
&&\qquad\qquad +\,(1+\tau^2)|\widetilde A_0(\ell)|^2,\nonumber\\
&& w_{\perp}(\ell) =
 \xi^2\,u\,\left(\widetilde A_1(\ell)|^2 -
 {\rm Re}[\widetilde A_0(\ell)\widetilde A_2^*(\ell)\right)
 -\tau^2|\widetilde A_0(\ell)|^2~,\nonumber\\
 \label{II3}
 \end{eqnarray}
 where $\tau^2=(u/u_\ell-1)\sin^2\phi_e$.
The basic functions $\widetilde A_m$ introduced in \cite{Titov2020}
are analogs of the well known IPA basis functions $A_m(n)$
in \cite{Ritus-79}:
 \begin{eqnarray}
 \widetilde A_m(\ell)
 =\frac{1}{2\pi}\int\limits_{-\infty}^{\infty}d\phi\,f^m(\phi)
 \cos^m(\phi+\tilde\phi)\,{\rm e}^{i\ell\phi
 -i{\cal P}^{(Lin)}(\phi)}
 \label{II4}
 \end{eqnarray}
with
\begin{eqnarray}
&&\qquad{\cal P}^{(lin)}(\phi) =
\tilde\alpha(\phi)- \tilde\beta(\phi)~,\label{II5}\\
&&\qquad\tilde\alpha(\phi)=\alpha\int\limits_{-\infty}^{\phi}d\phi'
f(\phi')\cos\phi'~,\label{II6}\\
&&\qquad\tilde\beta(\phi)=4\beta\int\limits_{-\infty}^{\phi}d\phi'
f^2(\phi')\cos^2\phi'~, \nonumber\\
&&\alpha=z\cos\phi_{e},\,\,
\beta=\frac{\xi^3u}{2\kappa},\,
\, z=\frac{4\xi^2u}{\kappa}\sqrt{\frac{u_\ell}{u}-1}~.
\label{II7}
\end{eqnarray}
The integrand of the function $\widetilde A_0(\ell)$ in Eq.~(\ref{II4})
does not contain the envelope function $f(\phi)$ and is, therefore, divergent.
It is regularized using the prescription of \cite{Boca-2009}
which leads to identity
\begin{eqnarray}
\ell\widetilde A_0(\ell) =  {\alpha}\widetilde A_1(\ell)
- 4\beta\widetilde A_2(\ell)~.
\label{II8}
\end{eqnarray}

The definitions in Eq.~(\ref{II3}) resemble corresponding IPA expressions,
i.e.\ for a monochromatic background field one has \cite{Ritus-79}
\begin{eqnarray}
&& w_{\parallel{n}} =
 \xi^2(u-1)\,(A_1^2 - A_0A_2)+\,(1+\tau^2)\,A_0^2,\nonumber\\
&& w_{\perp{n}} =
 \xi^2\,u\,(A_1^2 - A_0 A_2)
 -\tau^2\,A_0^2~,
 \label{II9}
 \end{eqnarray}
which can be obtained by replacing the basis functions
$\widetilde A_m(\ell)\to A_m\equiv A_m(n\alpha\beta)$
determined as
 \begin{equation}\label{II10}
 A_m(n\alpha\beta)=\frac{1}{2\pi}\int\limits_{-\pi}^{\pi} d\phi
 \cos^m(\phi)\,\,{\rm e}^{in\phi - i\alpha\sin\phi + i\beta\sin2\phi}
 \end{equation}
 with $z^2\to z^2(1+\xi^2/2)$, $\tau^2\to\tau^2(1+\xi^2/2)$,
 $\zeta\to\zeta(1+\xi^2/2)$,
 and with obvious substitutions
$\ell\to n$ in~(\ref{II1}). Thus, the corresponding
differential cross sections read
\begin{eqnarray}
\frac{d \sigma_i^{IPA}}{d\phi_{e} }
&=&\frac{2\alpha^2}{m^2\xi\kappa N_0}
\,\int\limits_{1}^{\infty}\frac{du}{u^{3/2}\sqrt{u-1}}
\,\sum_{n=n_{\rm min}}^\infty
\, {w_i}_n ,
\label{II11}
\end{eqnarray}
with $N_0=1/2$. The limit $n_{\rm min}$ is determined as through
integer part (Int) of
$n_{\rm min}={\rm Int}\left((4m^2(1+\xi^2/2))/s\right)+1$,
see \cite{Ritus-79,Titov2020} for details.

The cross section for an unpolarized incoming probe photon is
given by Eq.~(\ref{II1}) with
\begin{eqnarray}
w(\ell)=\frac{1}{2} (w_{\perp}(\ell) + w_{\parallel}(\ell))
\label{II12}
\end{eqnarray}
The difference in $w_{\perp}$ and $w_\parallel$ allows to introduce
the asymmetry for the total cross section, integrated over $\phi_e$, by
\begin{eqnarray}
{\cal A}=\frac{\sigma_{\perp}-\sigma_{\parallel}}
              {\sigma_{\perp}+\sigma_{\parallel}}~,
\label{II13}
\end{eqnarray}
as a function of $\zeta$, as well as the asymmetry of a function of $\phi_e$
at fixed $\zeta$ by
\begin{eqnarray}
{\cal A}(\phi_e)=\frac{d\sigma_{\perp}/d\phi_e-d\sigma_{\parallel}/d\phi_e}
                      {d\sigma_{\perp}/d\phi_e+d\sigma_{\parallel}/d\phi_e} .
\label{II14}
\end{eqnarray}

\section{Numerical results, \boldmath$\xi \leq 1$}

Below we present  our numerical results for the cross sections and asymmetries
for the the monochromatic laser beam (IPA) and for short
and ultra-short (sub-cycle) pulses (FPA).

\subsection{Infinite pulse (IPA)}

\begin{figure}[th]
\includegraphics[width=0.47\columnwidth]{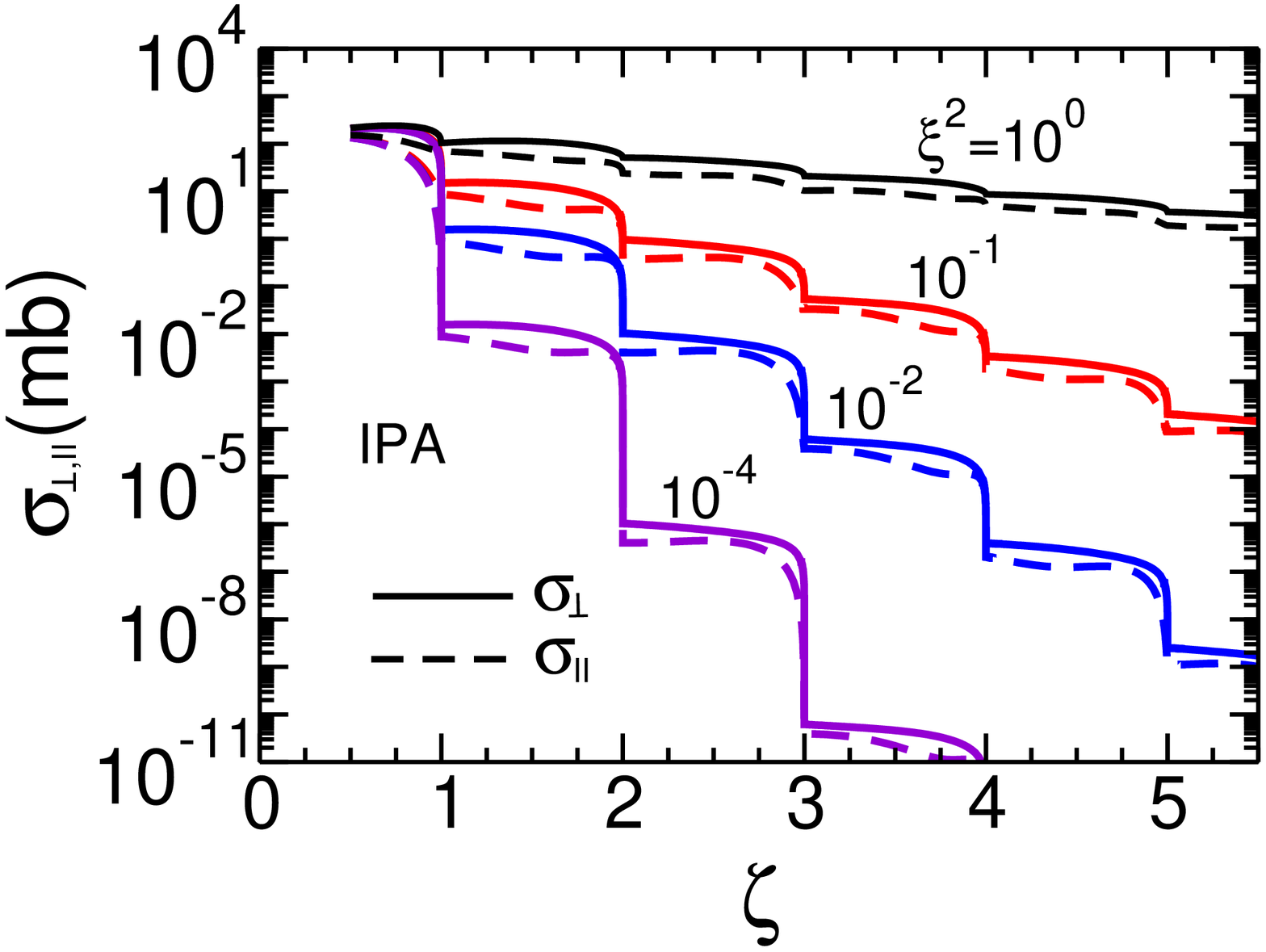} \hfill
\includegraphics[width=0.47\columnwidth]{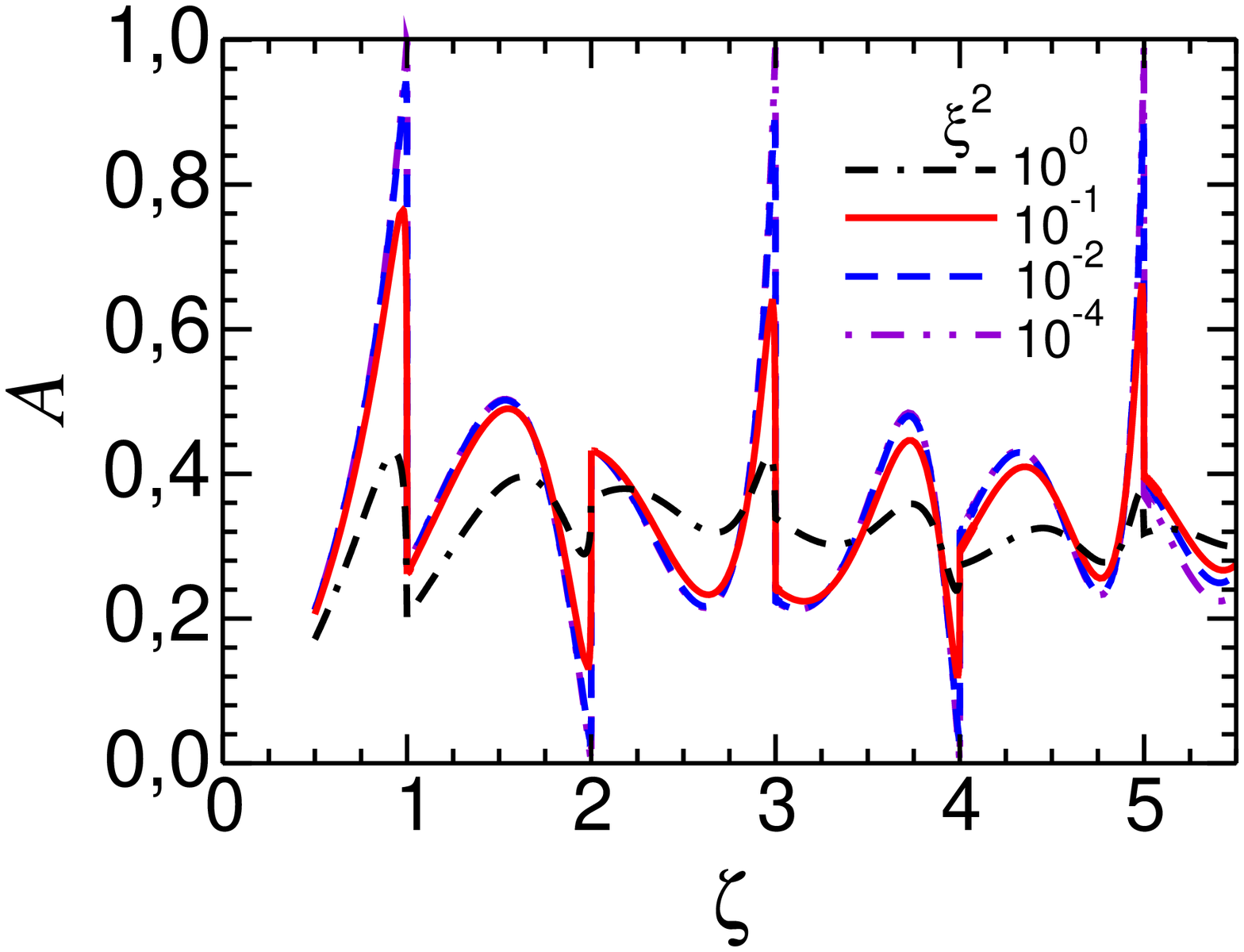}
\caption{\small{(Color online)
Results for a monochromatic laser beam, i.e.\
an infinitely long pulse (IPA).
Left panel: The cross sections $\spt$ (solid curves) and $\spl$ (dashed curves)
as a function of the threshold parameter $\zeta$ for
$\xi^2=10^0,\,10^{-1}$, $10^{-2}$ and $10^{-4}$.
Right panel: The asymmetry as a function of $\zeta$.
 \label{Fig:01}}}
 \end{figure}

The cross sections as a function of the
threshold parameter $\zeta$ are exhibited in the left panel of Fig.~\ref{Fig:01}
for $\xi^2=10^0,\,10^{-1},\,10^{-2}$ and $10^{-4}$. The cross sections
$\sigma_\perp$ and $\sigma_\parallel$ are depicted by solid and dashed
curves, respectively.

One can see a step-like behavior of the cross sections,
where each new step with $\zeta$ close to its integer
value $n_\zeta$ corresponds
to opening a new channel with the number of simultaneously
participating photons exceeding $n_\zeta$.
The step height is proportional to $\xi^{-2}$.
At $\xi^2=1$, the step-wise behavior practically
disappears, and the cross sections show an almost smooth
decrease with increasing $\zeta$.

 The asymmetry defined in Eq.~(\ref{II13}) as a function of the
 threshold parameter $\zeta$ is exhibited in Fig.~\ref{Fig:01}, right panel.
 For a weak field strength,
 $\xi^2\ll1$, the asymmetry exhibits sharp peaks and dips in the vicinity
 $\zeta\simeq 2m+1 -\epsilon$ and $\zeta\simeq 2m -\epsilon$
 with $m=0,1,2\dots$ and $\epsilon\ll1$, respectively.
 The height of the peaks (the depth of the dips) reaches
 a value of ${\cal A}\simeq1\,(0)$ at $\xi^2\le 10^{-4}$ and decreases
 (increases) with increasing values of $\xi^2$.

\begin{figure}[th]
\includegraphics[width=0.45\columnwidth]{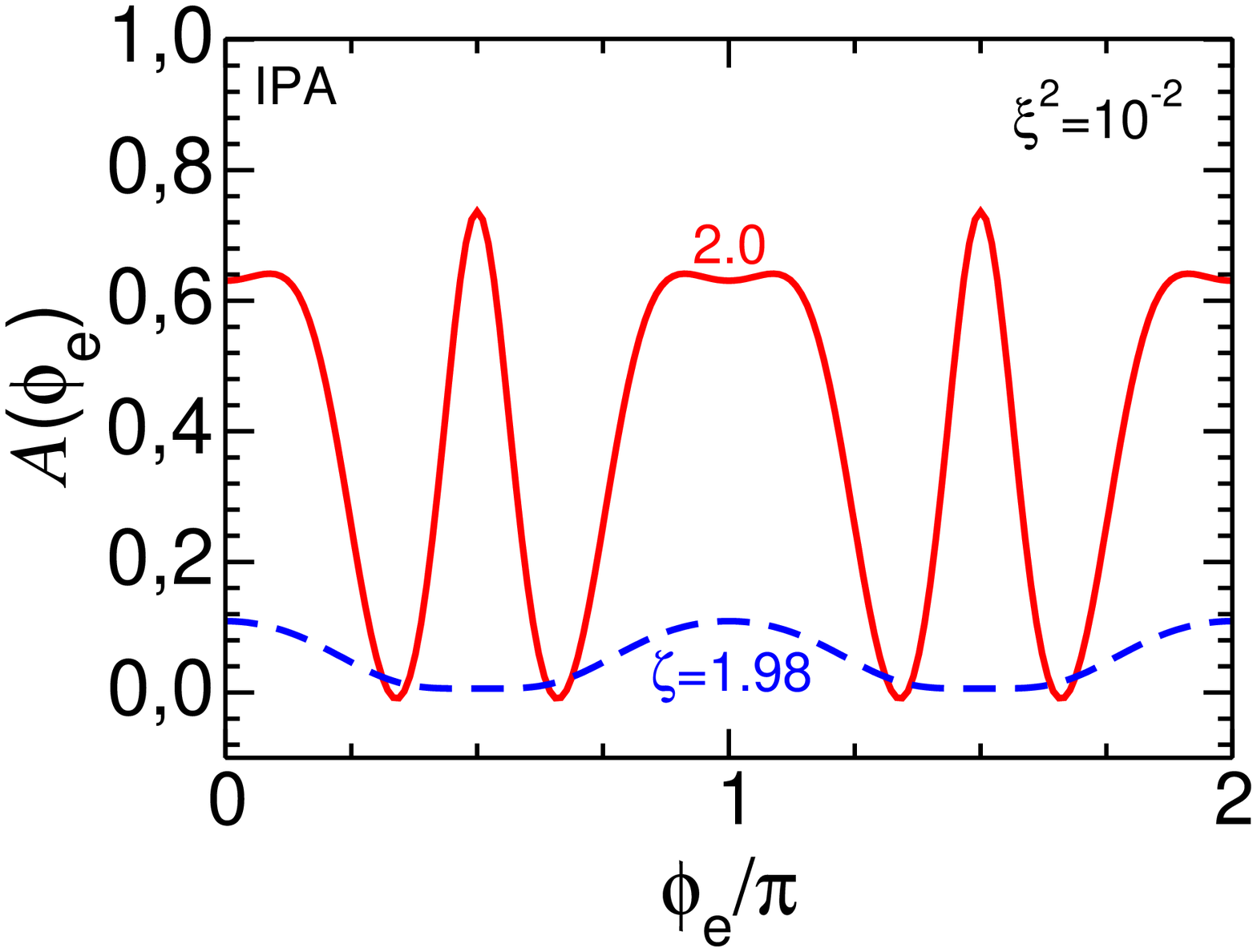} \hfill 
\includegraphics[width=0.45\columnwidth]{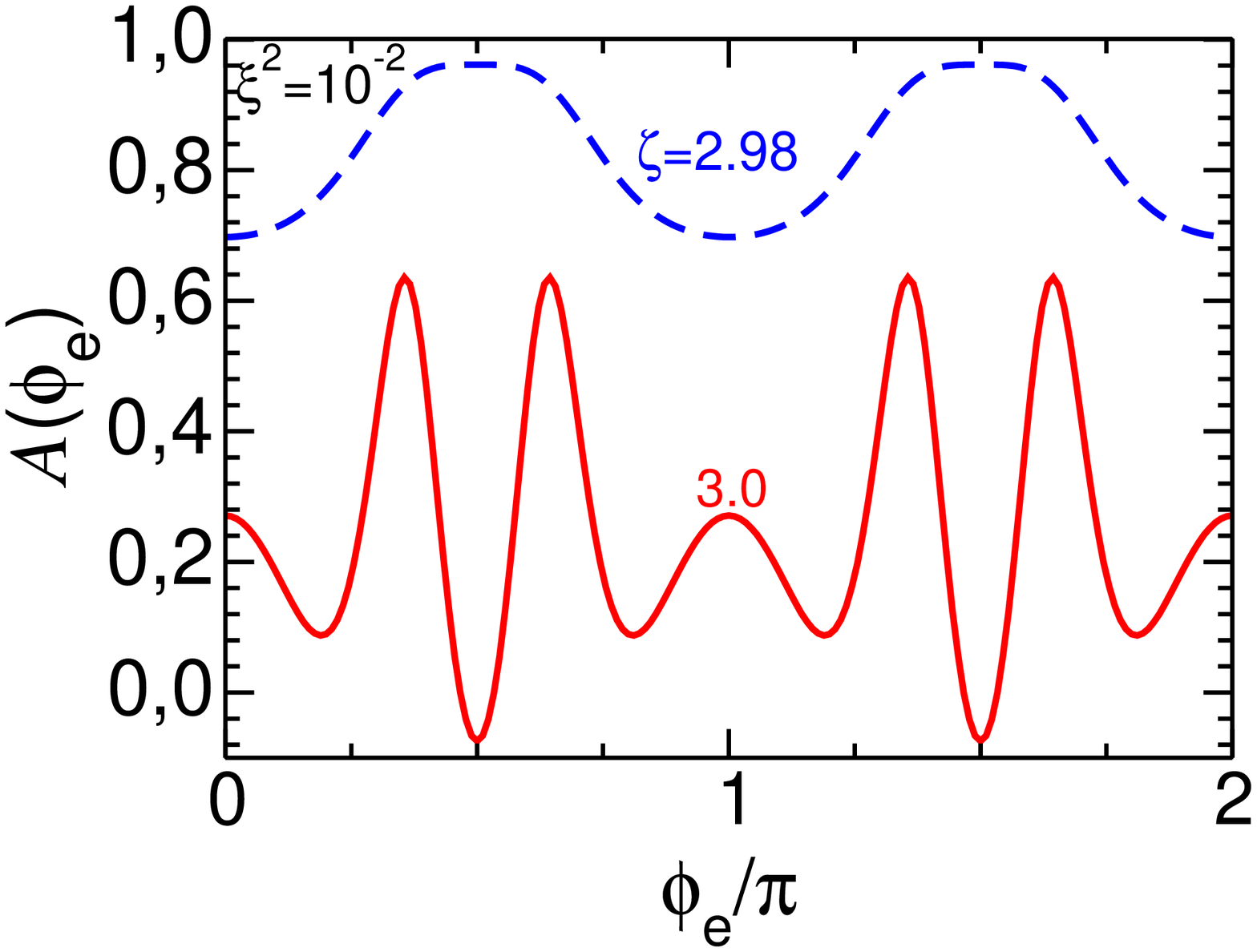}\\
\includegraphics[width=0.47\columnwidth]{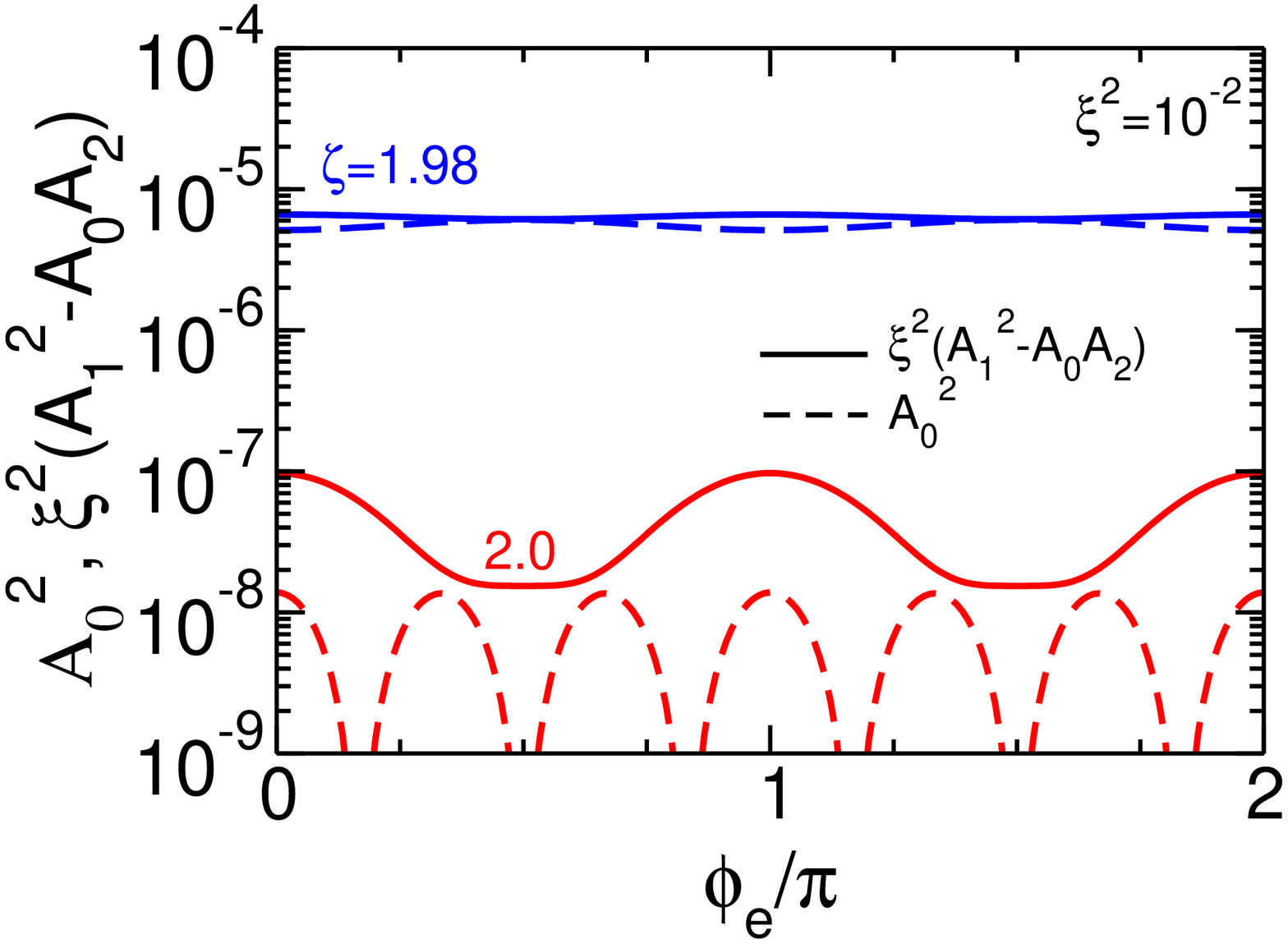} \hfill
\includegraphics[width=0.47\columnwidth]{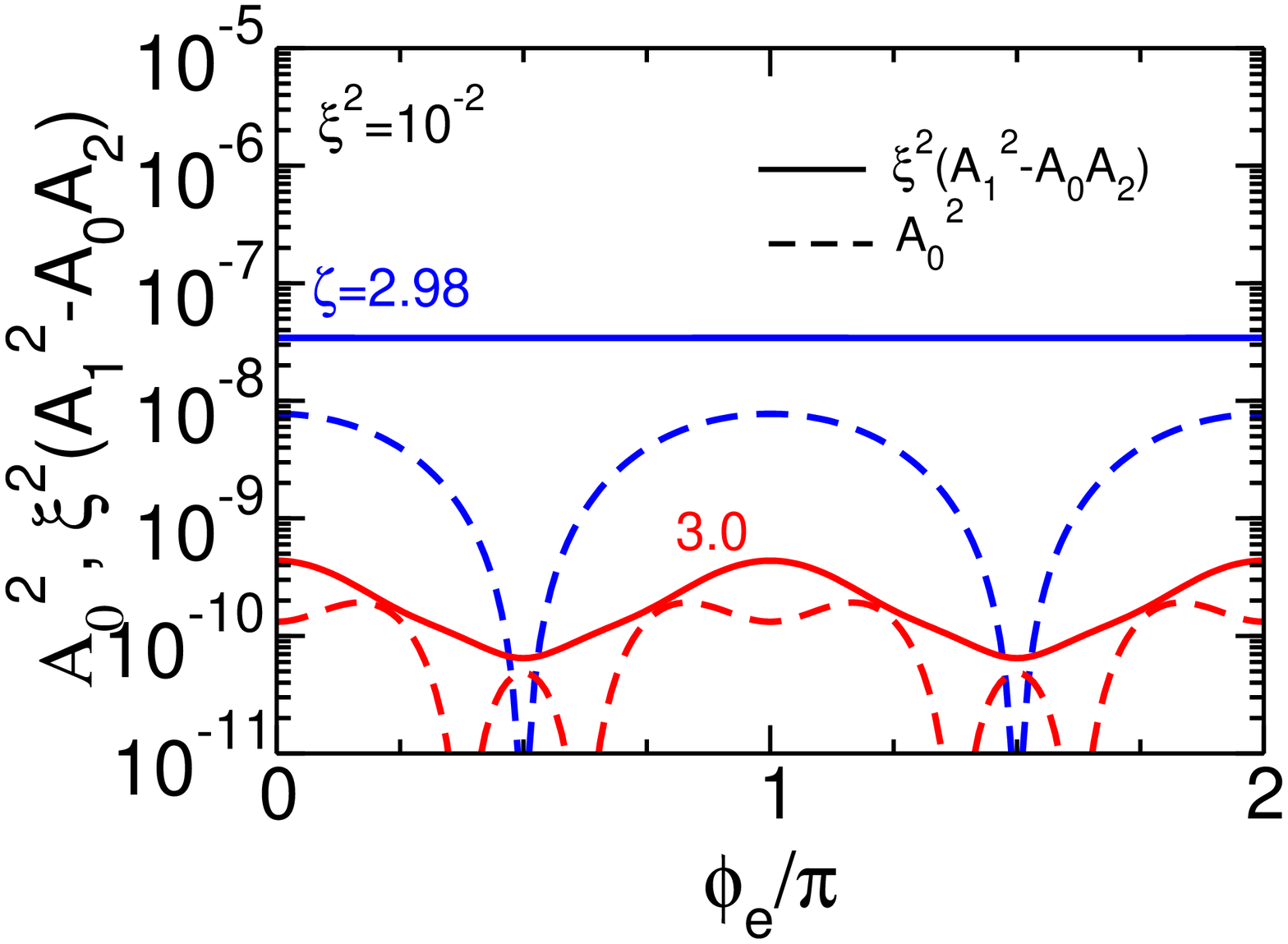}
\caption{\small{(Color online)
The asymmetry and combinations of $A_m$
as a function of $\phi_e$ for an infinitely long pulse (IPA) for $\xi^2=10^{-2}$.
The upper and top panels  are for asymmetries ${\cal A}$ and functions
$f_1=\xi^2(A_1^2-A_1A_2)$ and $f_2=A_0^2$, respectively;
the left and right panels correspond to $\zeta=2$ and 3,
respectively. The functions $f_1$ and $f_2$ are shown by solid and dashed
curves, respectively.
\label{Fig:03}}}
 \end{figure}

Sharp peak or dip positions correspond to the cases where
$\sigma_\perp\gg\sigma_\parallel$ or $\sigma_\perp\simeq\sigma_\parallel$,
respectively, and reflect the properties of the
basic functions $A_m(n)$.
Note that the non-monotonic $\zeta$ dependence of
asymmetry is determined by the numerator in the expression~(\ref{II13}).
In the here considered case of IPA one has
\begin{eqnarray}
&&{\cal A}(\phi_e)\propto\sum\limits_{n=n_{\rm min}}^{\infty}
\int\limits_{-1}^{1}v d\cos\theta_e\nonumber\\
&&\times\left(\xi^2\,(A_1^2 - A_0A_2)-(1+2\tau^2)\,A_0^2\right)\nonumber\\
&&\sim\xi^2\,(A_1^2 - A_0A_2)-\,A_0^2\mid_{n=n_{\rm min},\theta=\pi/2,\tau=0}~,
\label{II15}
\end{eqnarray}
neglecting a small, slowly varying variable $\tau^2$,
and taken integrand at  $\cos\theta=0$, $n=n_{\rm min}$
where the integral reaches its maximum value.

For an illustration, Fig.~\ref{Fig:03} exhibits
the asymmetry ${\cal A}(\phi_e)$ and the
combination of the functions $f_1=\xi^2(A_1^2 - A_0A_2)$ and $f_2=A_0^2$,
which are depicted in the top and bottom panels, respectively.
The dependence on the azimuthal angle in the vicinity of $\zeta=$2 and 3,
is shown in the left and right panels, respectively.
When $\zeta=2-\epsilon$, then $f_1\simeq f_2$, which leads to a small
asymmetry and manifests itself in a dip in the right panel of Fig.~\ref{Fig:01}.
For $\zeta=3-\epsilon$, one gets $f_1\gg f_2$ and, as a result, the asymmetry
exhibits a sharp peak in the right panel of Fig.~\ref{Fig:01}.

Finally, we conclude that in the region under consideration,
$\xi^2\le1 $ and $\zeta\le5 $, the asymmetry shows sharp peaks and dips
in such a way that for $\xi^2\le10^{-4}$ the asymmetry
varies in the range $0\lesssim{\cal A}\lesssim 1$.
With increasing values of $\xi^2$, the range of
the variation decreases significantly.
At $\xi^2=1$, the asymmetry varies in  the range $0.2\dots0.4$
with average value $\approx 0.33$.

\subsection{Finite pulse (FPA)}

\begin{figure}[th]
\includegraphics[width=0.5\columnwidth]{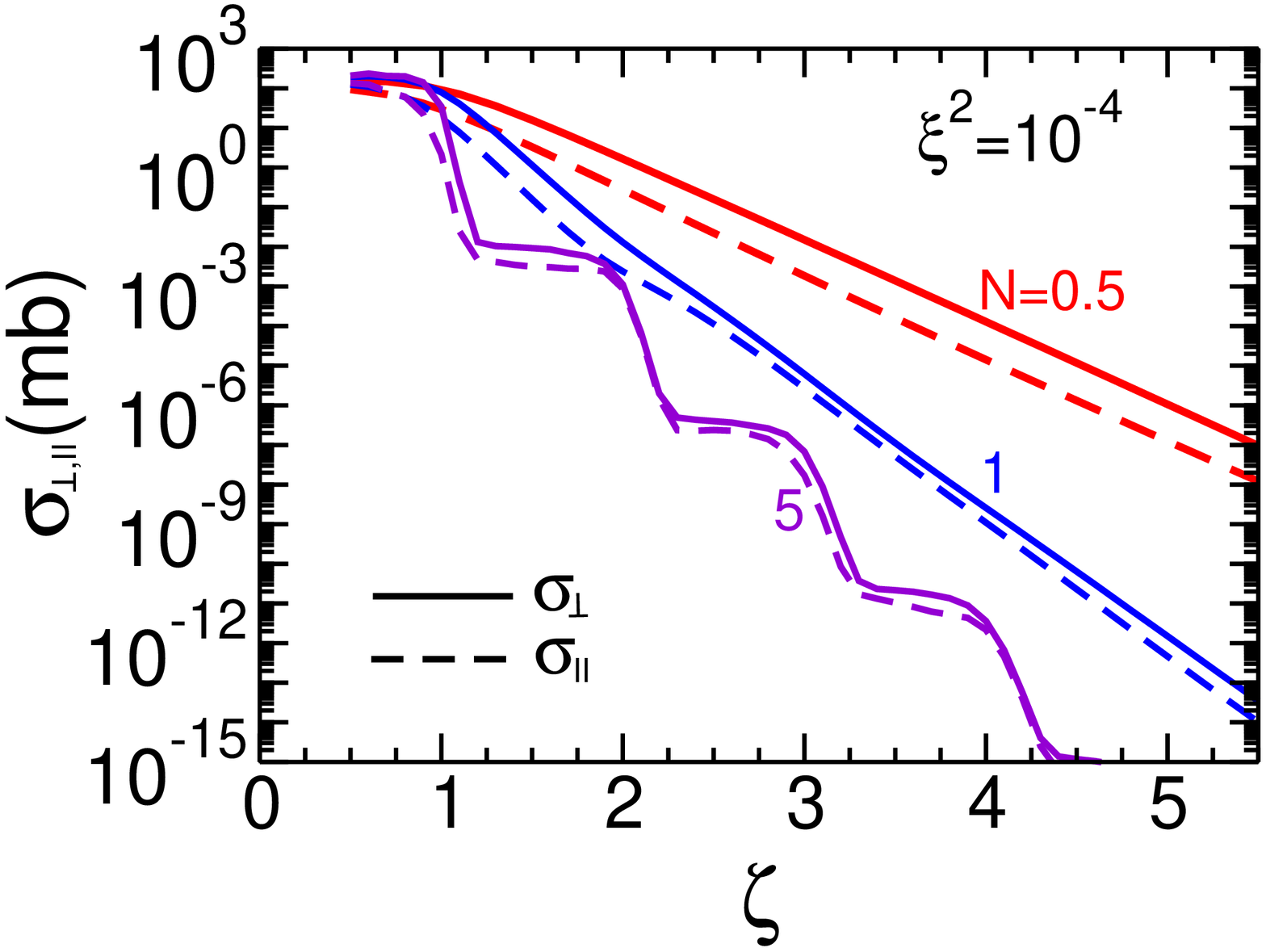} \hfill
\includegraphics[width=0.475\columnwidth]{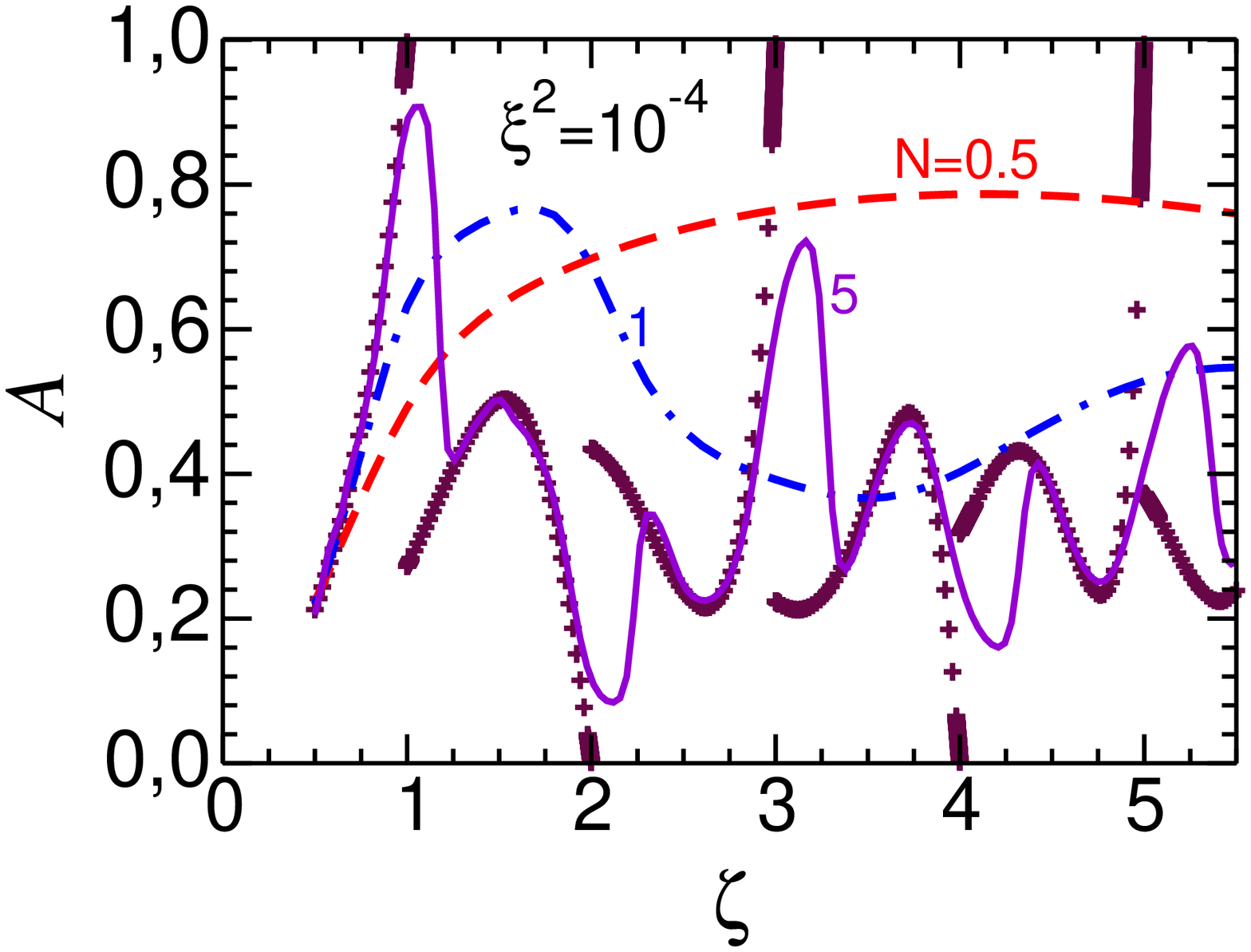}\\
\includegraphics[width=0.5\columnwidth]{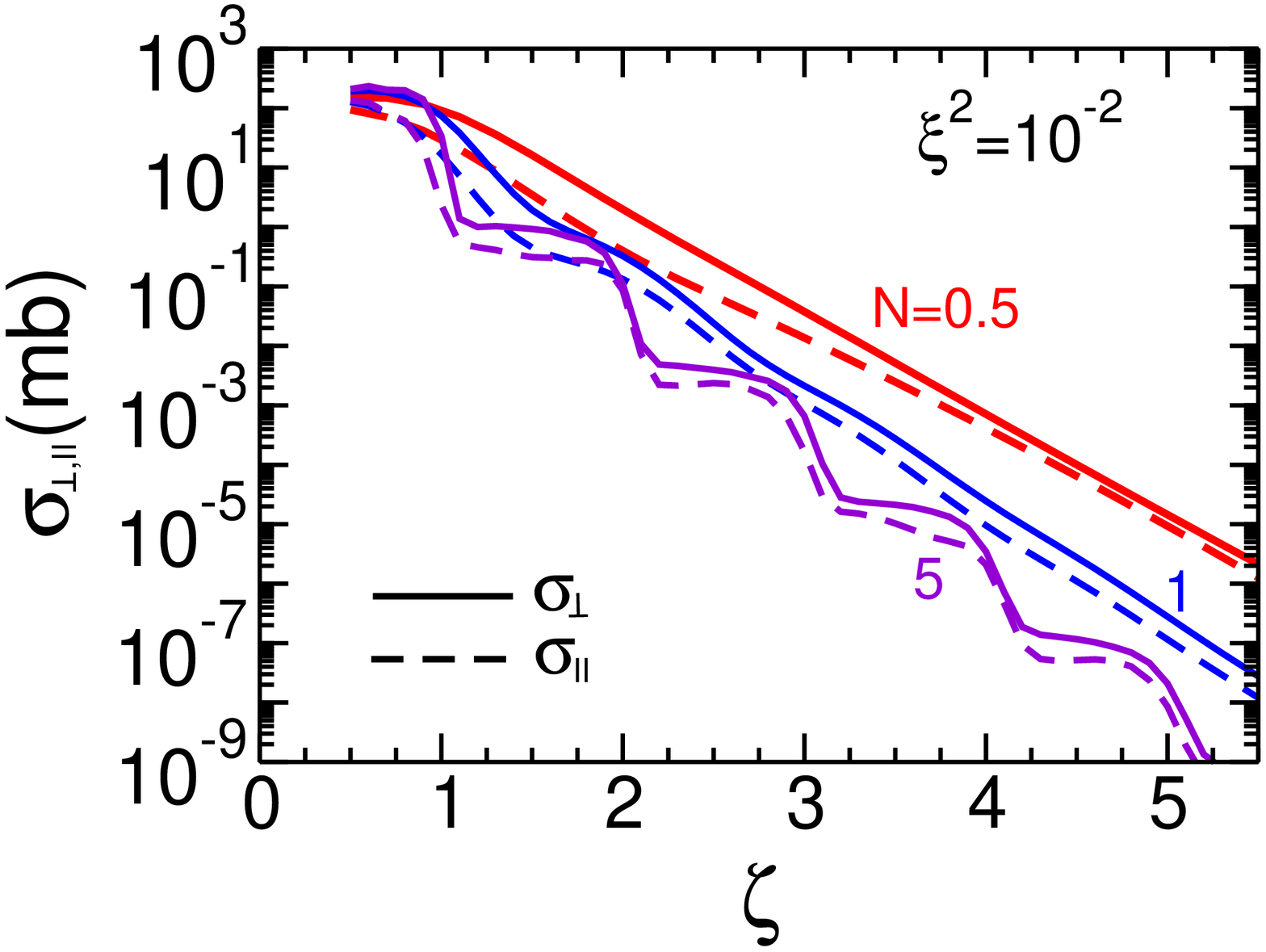} \hfill
\includegraphics[width=0.475\columnwidth]{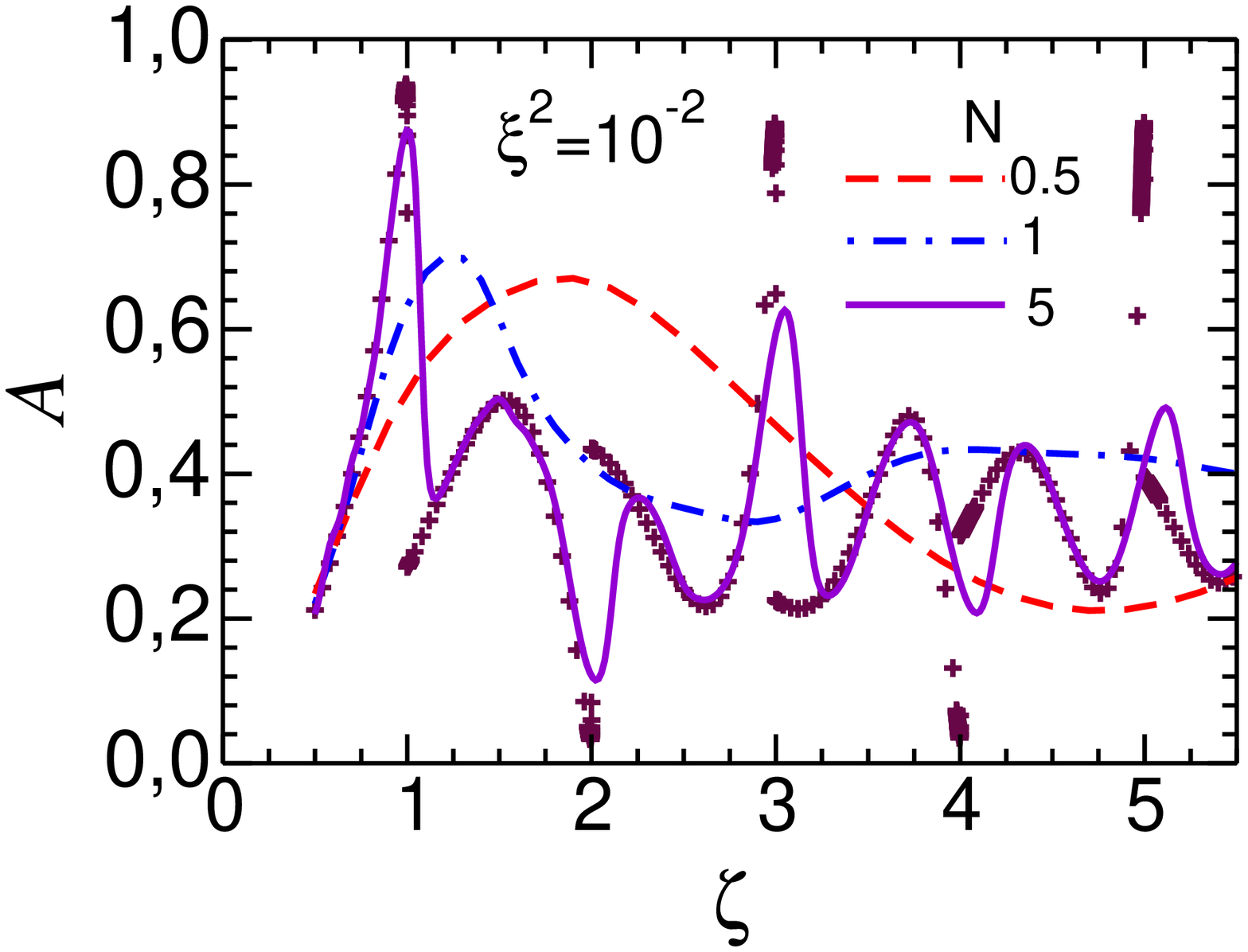}\\
\includegraphics[width=0.5\columnwidth]{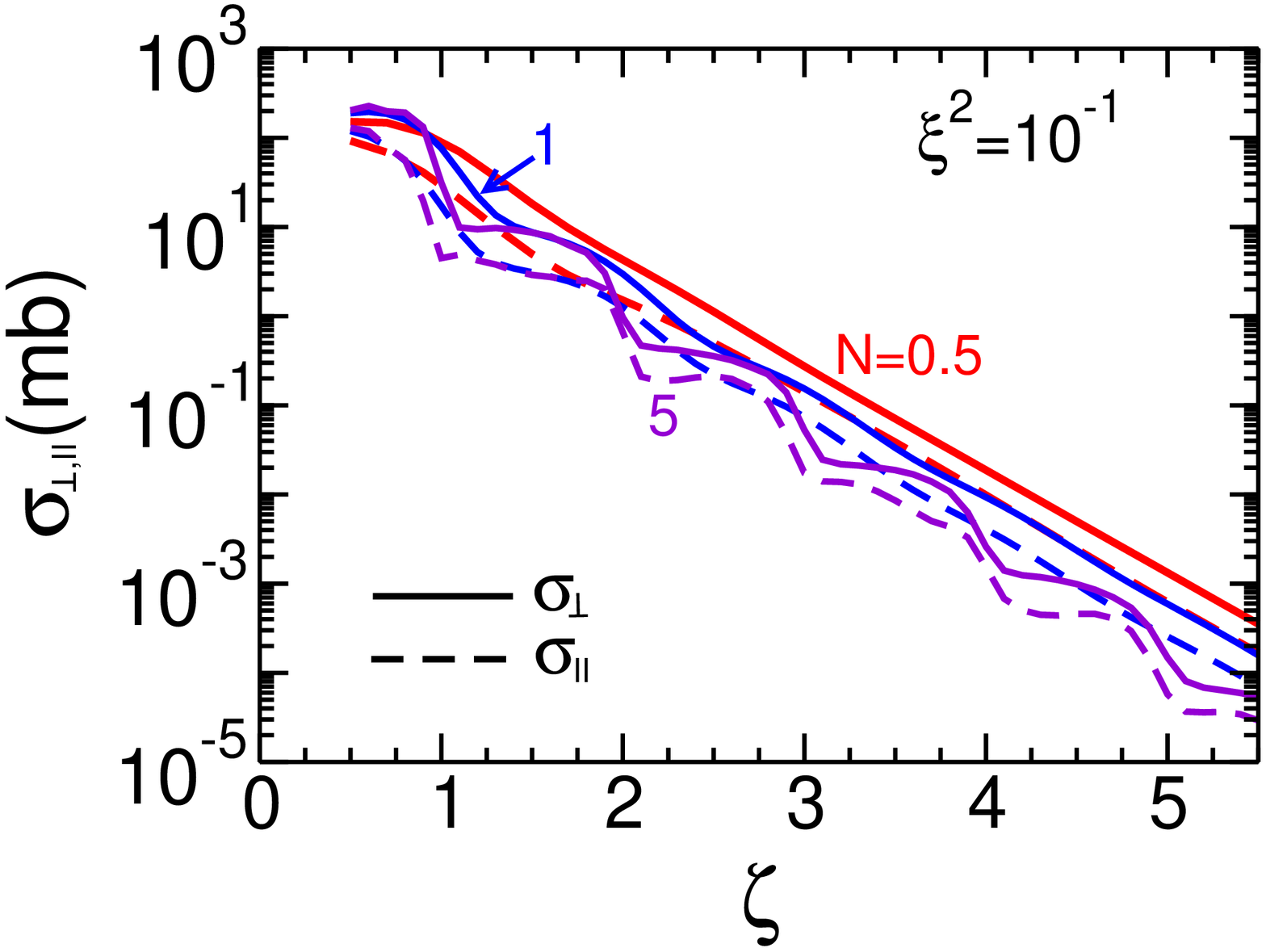} \hfill
\includegraphics[width=0.475\columnwidth]{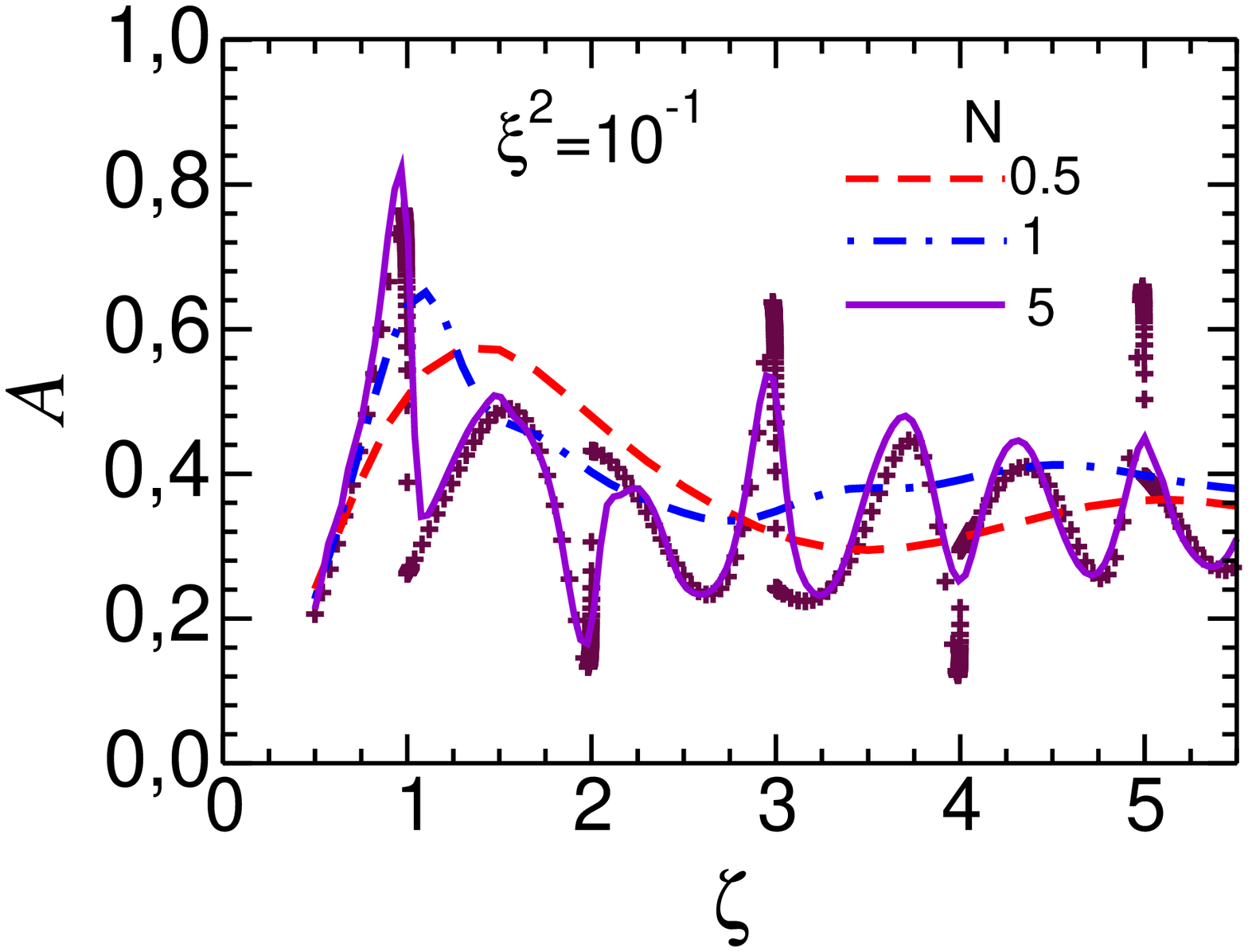}\\
\includegraphics[width=0.5\columnwidth]{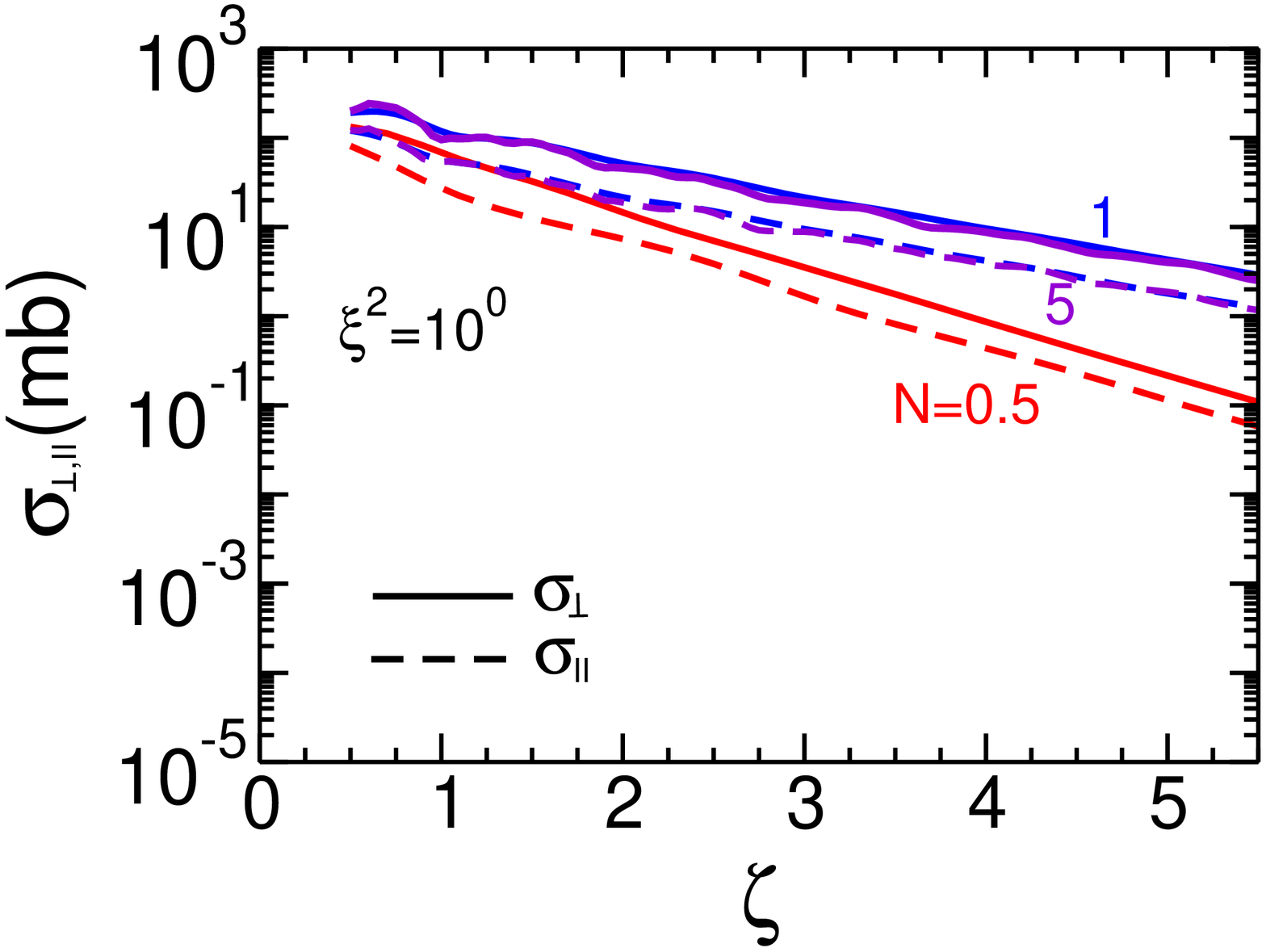} \hfill
\includegraphics[width=0.475\columnwidth]{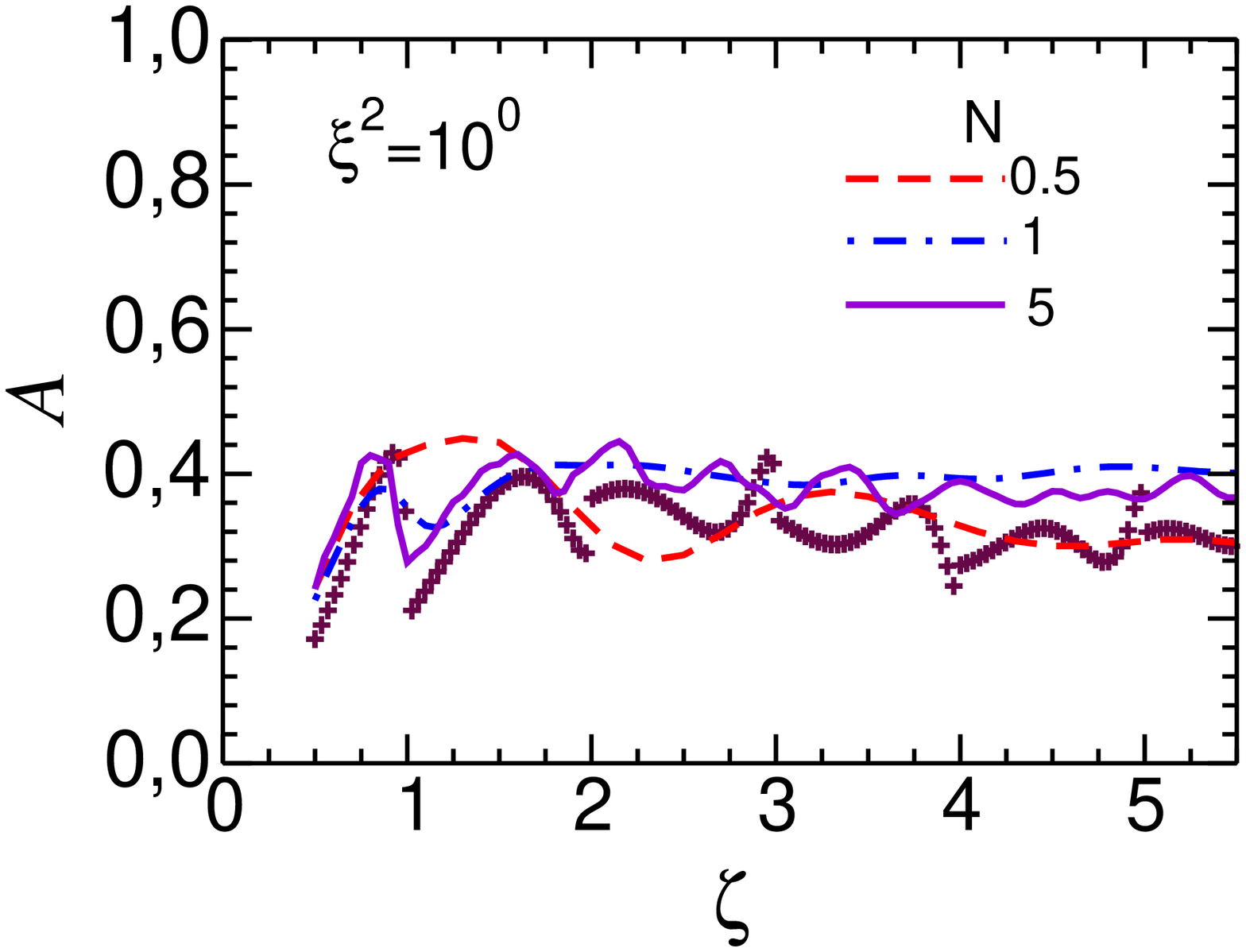}
\caption{\small{(Color online)
Left panels: The total cross sections of the non-linear BW
$\ee$ pair production as a function of the threshold parameter $\zeta$
for finite pulses with $N=$0.5, 1 and 5. The results for
$\xi^2=10^{-4}$, $10^{-2}$, $10^{-1},$ and $10^{0}$
are displayed sequentially from the top to bottom panels.
The solid and dashed curves are for $\sigma_{\perp}$ and $\sigma_{\parallel}$, respectively.
Right panels: The asymmetry as a function of $\zeta$
for different values of $\xi^2$ and $N$.
The dashed, dot-dashed and solid curves are for $N=0.5$, 1 and 5, respectively.
Crosses are for the infinite pulse (IPA) prediction.
 \label{Fig:04}}}
 \end{figure}

Our results for non-linear Breit-Wheeler $\ee$ pair production
as a function of the threshold parameter $\zeta$ for finite pulses
(FPA) with different pulse lengths (characterized by $N$)
and field intensity $\xi^2$ are exhibited in Fig.~\ref{Fig:04}.

The total cross sections $\spt$ and $\spl$ as a function of $\zeta$
for $\xi^2=10^{-4}$, $10^{-2}$, $10^{-1},$ and $10^{0}$
are displayed sequentially from the top to bottom panels
in Fig.~\ref{Fig:04}~(left).
The solid and dashed curves are for  $\sigma_{\perp}$
and $\sigma_{\parallel}$, respectively.
In the case of $N=5$ and $\xi^2\lesssim 10^{-1}$,
the cross sections exhibit
a step-like structure with steps near the integer values
of $\zeta$, similar to the IPA prediction shown in Fig.~\ref{Fig:01}.
The height of the steps $\propto \xi^{-2}$.
At $\xi^2=1$, the step-wise structure of the cross sections
goes into an almost monotonic decrease with increasing $\zeta$.
The cross sections for $N=1$ and 5 are close to each other,
i.e.\ the result becomes insensitive to the pulse duration.

In the cases of short ($N=1$) and very short (sub-cycle, $N=0.5$) pulses,
the cross sections exhibit a monotonic exponential decrease
with increasing $\zeta$.

The difference between $\sigma_{\perp}$ and $\sigma_{\parallel}$
leads to a finite asymmetry, displayed in the right panels of
Fig.~\ref{Fig:04}. For convenience, the prediction for the IPA case
is shown by crosses.
Consider first the case of a weak field, i.e.\ an intensity
referring to $\xi^2\lesssim10^{-1}$.
At a relatively large pulse width with $N \ge 5$, the asymmetry resembles
qualitatively IPA result (cf.\ Fig.~\ref{Fig:01}, right)
with some peaks and dips. Their positions are close to that of the IPA case.

At sub-cycle  and short pulses with $N=0.5$ and 1,
respectively, the asymmetries exhibit smooth non-monotonic behavior without
sharp peaks and dips.

Let us analyze the asymmetry as a function of the azimuthal angle $\phi_e$.
First, we consider the case of $\zeta$ being in the interval
between the two nearest integer values, i.e.\ away from
the values for which IPA predicts sharp peaks/dips.
\begin{figure}[th]
\includegraphics[width=0.45\columnwidth]{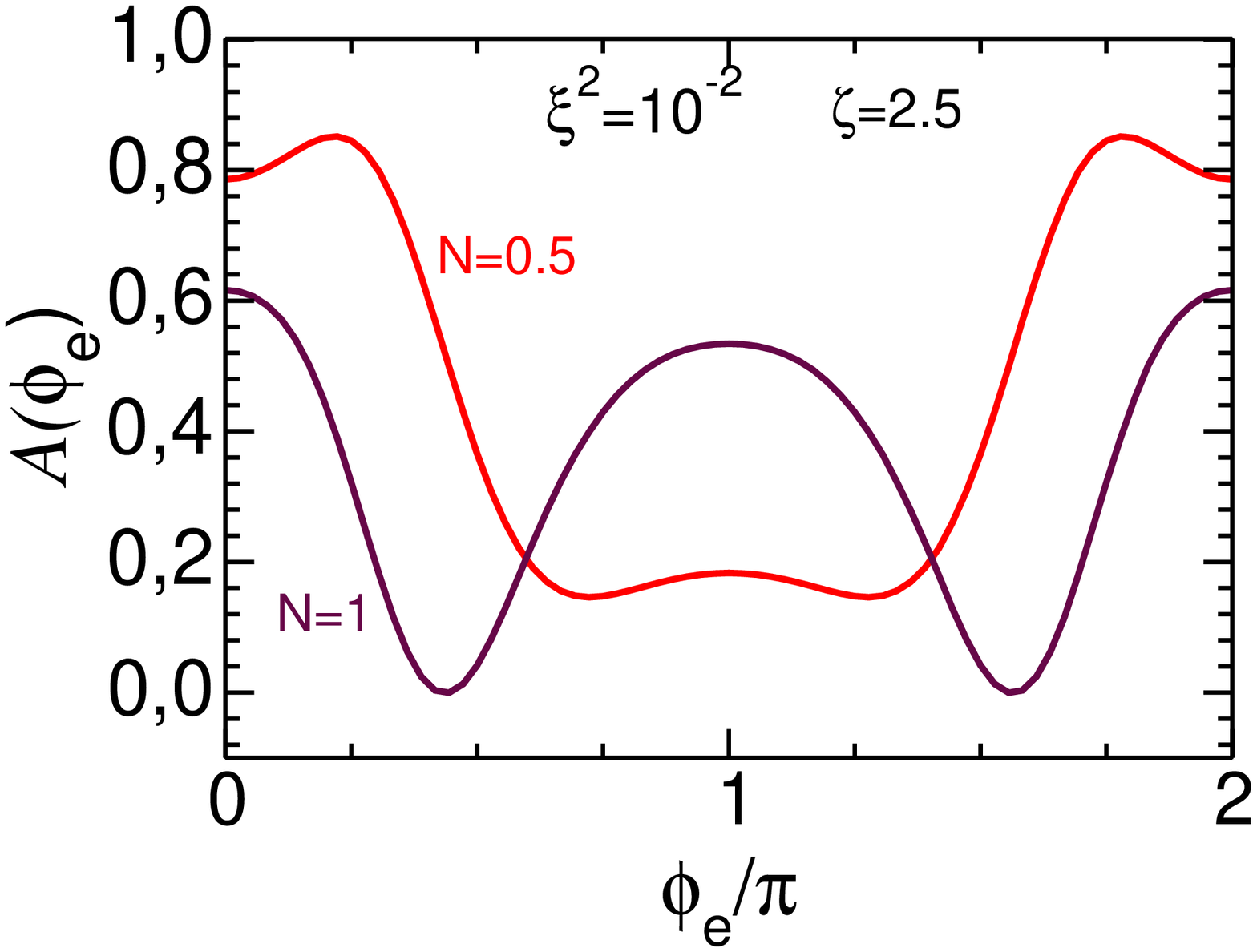} \hfill
\includegraphics[width=0.45\columnwidth]{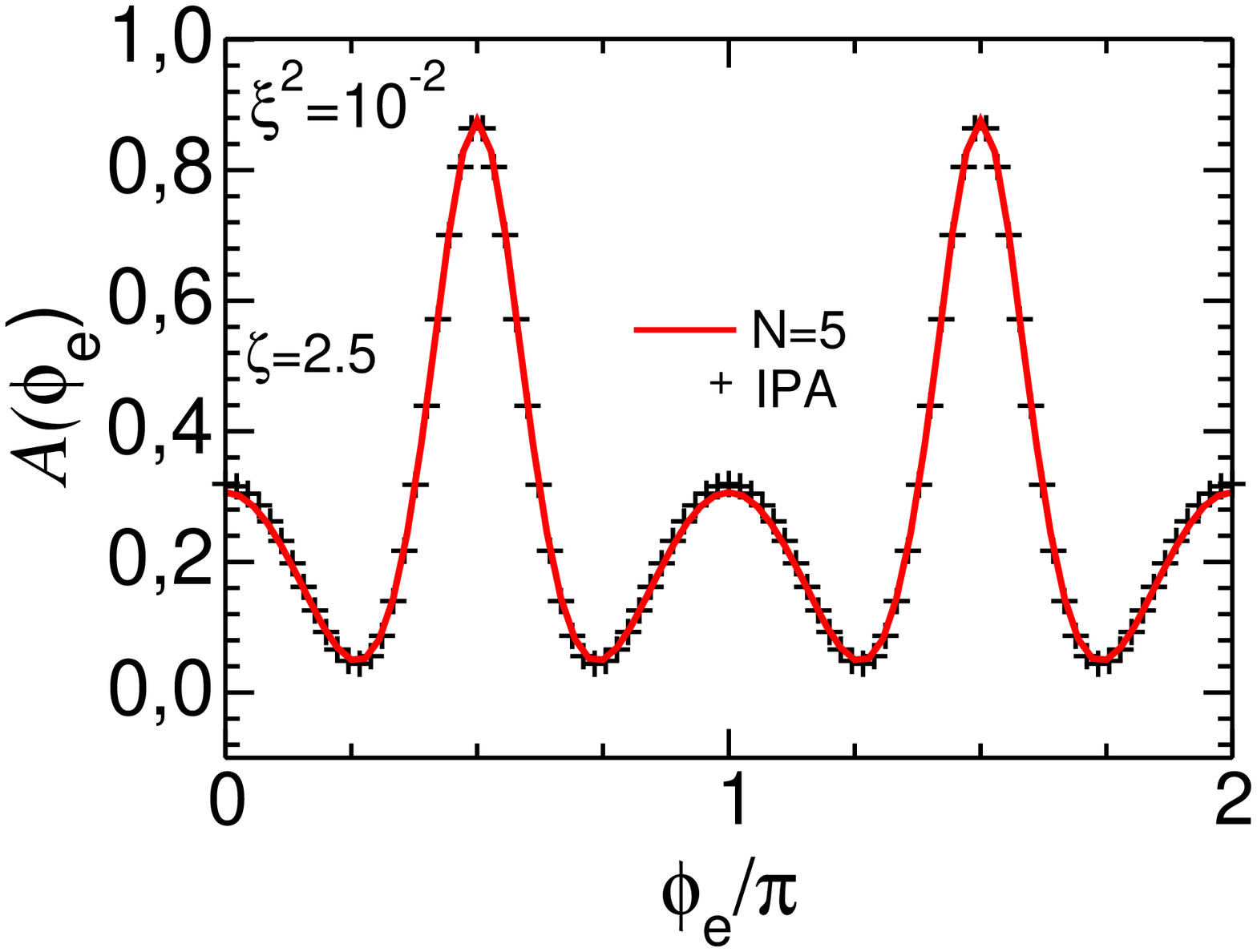}
 \caption{\small{(Color online)
The asymmetry as a function of $\phi_e$ for finite pulses
characterized by $N=0.5$, 1 (left panel) and $N=5$ (right panel) at
$\zeta=$2.5  and $\xi^2=10^{-2}$.
The crosses in the right panel are for the IPA case.
 \label{Fig:05}}}
 \end{figure}

Our result for $\zeta=2.5$ and $\xi^2 = 10^{-2}$
is depicted in Fig~\ref{Fig:05}. Predictions for very short
pulses with $N\le1$ and short pulses with $N\ge5$ are different,
therefore, they are shown separately.

For pulse with $N=0.5$, the asymmetry is quite a large at
$\phi_e=0$ (${\cal A}(0)\simeq 0.8$),
increases toward a local maximum at $\phi\simeq\pi/4$,
and then decreases toward a minimum at $\phi_e=\pi/2$.
(Note that the asymmetry is symmetric under the substitution
$\phi_e\to2\pi-\phi_e$.)
For the short pulse with $N=1$, the asymmetry has a maximum at $\phi_e=0$,
then it decreases up to zero at $\phi_e/\pi\simeq 0.44$ and has a maximum at
$\phi_e=\pi$.

In case of a pulse with $N=5$, the asymmetry exhibits local maxima at
$\phi_e=0\,\pi/2,\,\pi$ and minima at $\pi/4,\,3\pi/4$.
The result for the finite pulse coincides practically
with the prediction for the IPA case, shown by crosses.
\begin{figure}[th]
\includegraphics[width=0.47\columnwidth]{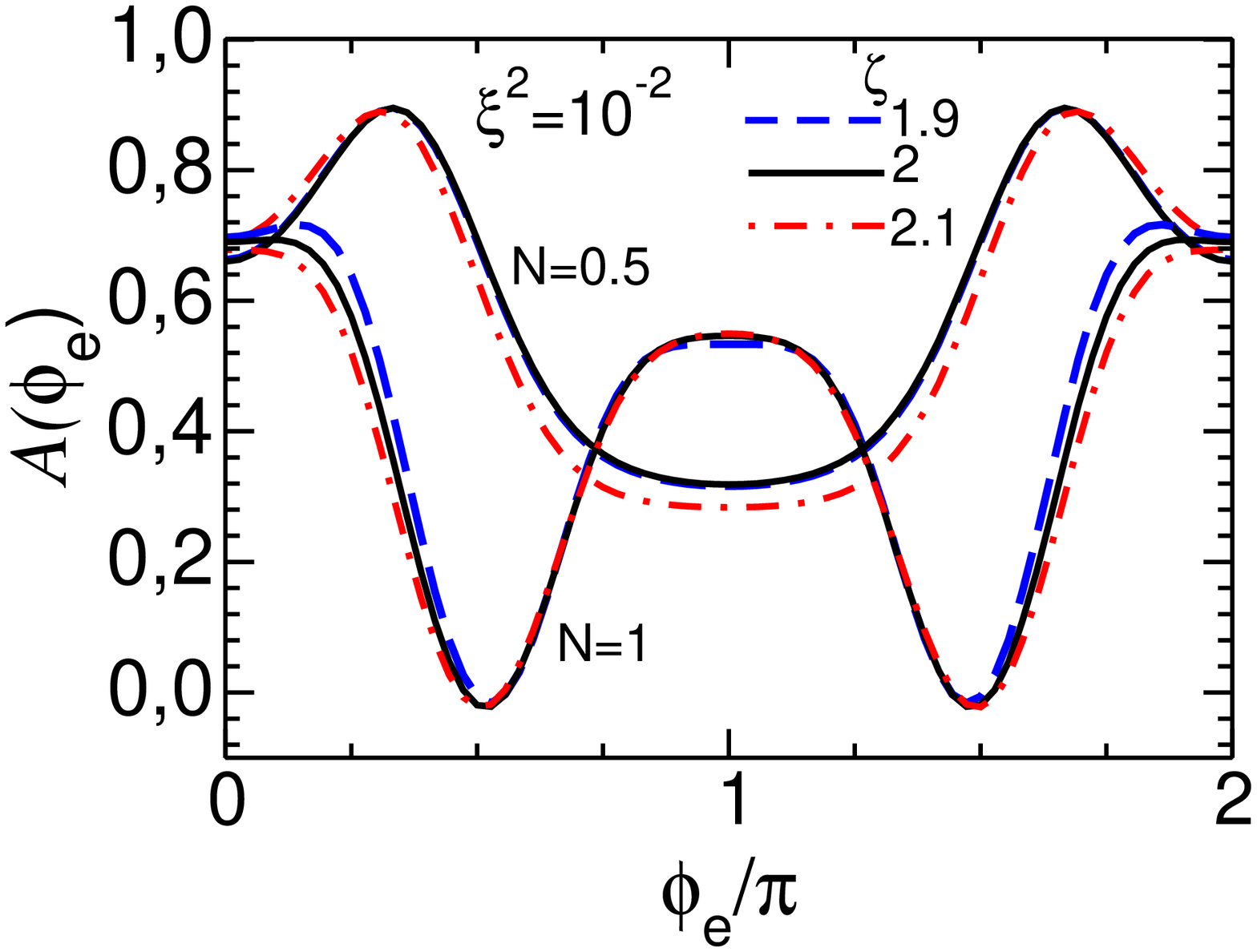} \hfill
\includegraphics[width=0.47\columnwidth]{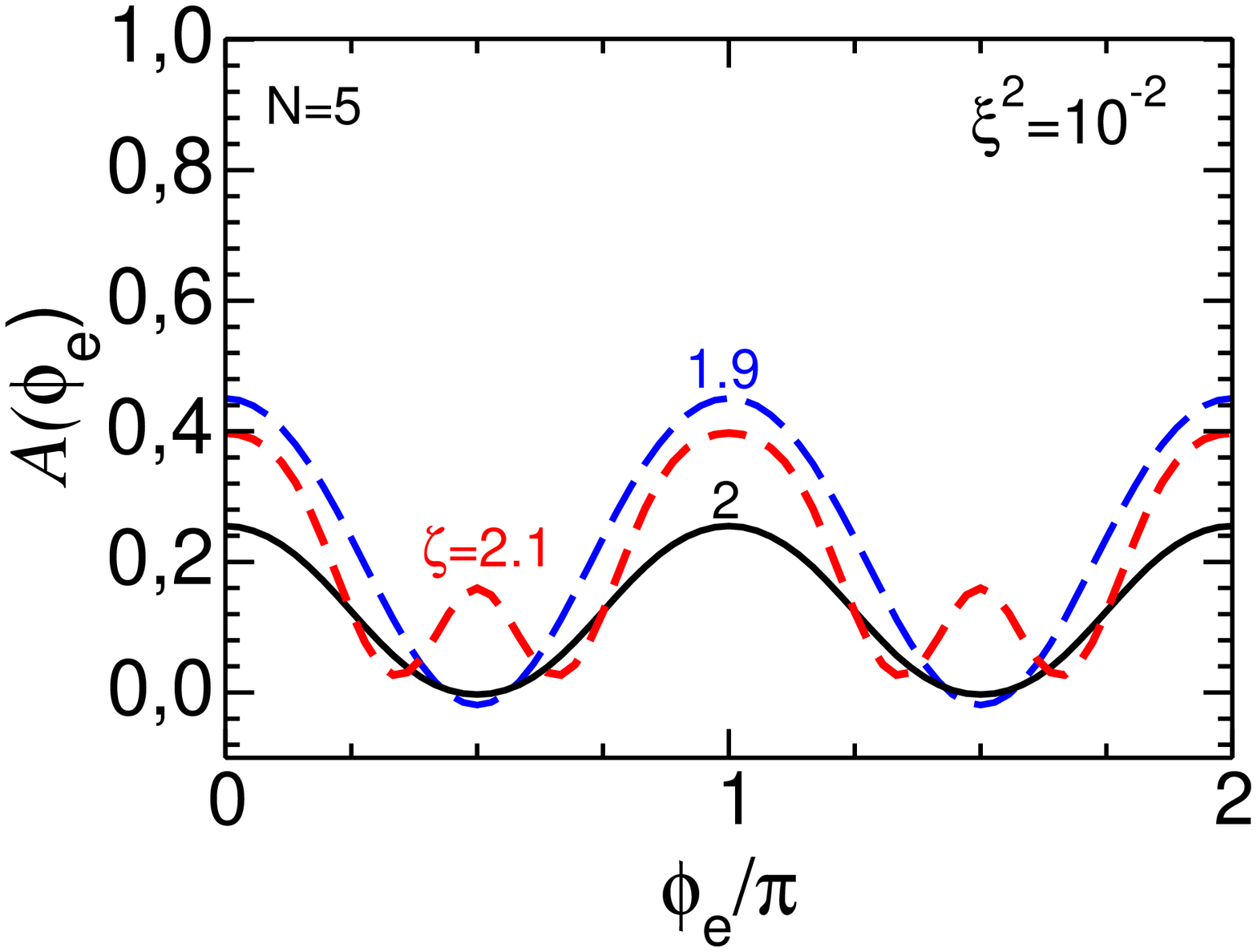}
\caption{\small{(Color online)
The asymmetry as a function of $\phi_e$ for finite pulses
with $N=0.5$, 1 (left panels) and $N=5$ (right panels) and
$\zeta=$1.9, 2, 2.1 at $\xi^2=10^{-2}$.
 \label{Fig:06}}}
 \end{figure}

For completeness,  in Fig.~\ref{Fig:06} we also present
results for the asymmetry in the vicinity of integer values of $\zeta=2$.
The asymmetries for $N=0.5$, 1 and $N=5$ are shown in the left and right panels,
respectively. In our calculations,
we choose $\zeta=1.9$, 2, and 2.1 at $\xi^2=10^{-2}$.
In the case of short pulses with $N\le1$, the result is similar to that shown
in  Fig.~\ref{Fig:06} (left panel) and, within the chosen interval, is practically
independent of $\zeta$. For $N\ge 5$,  the model predicts clear maxima
at $\phi_e=0,\,\pi$.  An additional bump occurs at $\phi_e=\pi/2$ and
$\zeta=2.1$, but its height is much smaller than that predicted for
$\zeta=2.5$ (cf.\ Fig.~\ref{Fig:05} (right panel)).

\section{Cross sections and asymmetry in ultra-intense fields, \boldmath$\xi \gg 1$}

At large values of $\xi\gg1$, the main contribution to the $\ee$ pair production
comes from the central part of envelope, the final result is not sensitive
to the envelope structure \cite{TitovPRA}, and for calculation of the
partial probabilities one can use the IPA formalism
with replacing the summation $\sum\limits_n$ in Eq.~(\ref{II11})
by the integration $\int d\ell$.
The total cross section reads
\begin{eqnarray}
\frac{d \sigma_{i}}{d\phi_{e} }
=\frac{2\alpha^2}{m^2\xi\kappa N_0}
\,\int\limits_{0}^{2\pi} d\phi
\,\int\limits_{1}^{\infty}\frac{du}{u^{3/2}\sqrt{u-1}}
\,\int\limits_{\ell_{\rm min}}^{\infty}d\ell\,{w_{i}}_{\ell}~,
\label{III1}
\end{eqnarray}
where $\phi=\phi_e$ and
$\ell_{\rm min}=4m^2(1+\xi^2/2)u/s$,
and $N_0=1/2$.
The corresponding formalism was developed by Nikishov and
Ritus~\cite{RitusGroup,Ritus-79}.
For completeness and easy reference, we provide in this section
the most important expressions of their approach,
necessary for the subsequent analysis.

The continuous variable $\ell$ is expressed in terms of
the auxiliary variables $\rho$ and $\tau$ via
\begin{eqnarray}
\ell=\frac{2\xi^2u}{\kappa}
\left(\rho^2 +\frac{\tau^2}{\xi^2}\right) +\ell_{\rm min} ,
\label{III2}
\end{eqnarray}
where
\begin{eqnarray}
\rho^2&=&\frac{1}{\xi^2}\left(1+\frac12\xi^2\right)
\left(\frac{u_\ell}{u}-1\right)\cos^2\phi~,
\nonumber\\
\tau^2&=&\left(1+\frac12\xi^2\right)
\left(\frac{u_\ell}{u}-1\right)\sin^2\phi~,
\label{III3}
\end{eqnarray}
with $\ell_{\rm min}\equiv\ell_0=
2m^2(1+\frac12\xi^2)/(k\cdot k')
=2\xi(1+\xi^2/2)/\kappa$ and $u_\ell=\ell/\ell_0$.
The variables $\rho$, $\tau$ and $\phi$ allow to perform
a useful transformation
\begin{eqnarray}
\int\limits_{0}^{2\pi}\,d\phi\int\limits_{\ell_{\rm min}}^{\infty}\,d\ell=\frac{4\xi^2u}{\kappa}
\int\limits_{-\infty}^{\infty}\,d\rho\int\limits_{-\infty}^{\infty}\,d\tau .
\label{III4}
\end{eqnarray}
Further, for large $\ell$ the arguments $\alpha$ and $\beta$ in
the basic functions $ A_m(\ell\alpha\beta)$ in Eq.~(\ref {II10})
are also large and, therefore, the bi-linear combinations of $A_0^2$
and $A_1^2-A_0A_2$ in (\ref{II9}) can be replaced
by asymptotic expressions:
\begin{eqnarray}
&&A_0^2=\frac{2\sigma}{\pi^2\xi^2y\,\sin^2\psi}
\Phi^2(y)(1+\cos2\eta)~\nonumber\\
&&A_1^2-A_0A_2=\frac{2\sigma^2}{\pi^2\xi^4y^2\,\sin^2\psi}
\left(y\,\Phi^2(y)+\Phi'{}^2(y)\right.
\label{III5}\\
&&\left.+(y\,\Phi^2(y)+\Phi'{}^2(y))\cos2\eta\right)~,
\label{III6}
\end{eqnarray}
where $\Phi(y)$ and $\Phi'(y)$ denote the Airy function and its derivative,
respectively. The variables $\psi,\,\sigma$, and $\eta$ are associated with
the variables $\ell,\,u,\,\rho,$ and $\tau$ as
\begin{eqnarray}
&&\cos\psi=\rho,\,\,\sigma=1+\tau^2,\,y=\sigma
\left(\frac{2u}{\kappa\sin\psi}\right)^{\frac 23},\nonumber\\
&&\eta=\ell\left(\psi -\frac{\sin2\psi}{2(1+2\cos^2\psi)} \right) .
\label{III7}
\end{eqnarray}
By making use of Eqs.~(\ref{III5}) - (\ref{III7}) and discarding highly-oscillating
terms proportional to $\cos2\eta$, one can obtain the final expressions
for $\sigma_{i}$ ($i={\perp},\,{\parallel}$)
\begin{eqnarray}
\sigma_{i\,\infty}=
\frac{32\alpha^2}{\pi^2\,\xi\kappa\,m^2}
\int\limits_0^\pi\,d\psi
\int\limits_0^\infty
\frac{dt}{(t^2+1)^\frac32}
\int\limits_{-\infty}^{\infty}d\tau
w_i
\label{III8}
\end{eqnarray}
with $t=\sqrt{u-1}$ and
\begin{eqnarray}
&&w_{\perp\,\infty}=a\left(u\,F_1 -\tau^2F_2\right)~,\nonumber\\
&&w_{\parallel\,\infty}=a\left((u-1)\,F_1+ (1+\tau^2)F_2\right)~,
\label{III9}\\
&&F_1=b\,(y\,\Phi^2(y) + \Phi(y)'{}^2),\,F_2=\Phi^2(y),\nonumber\\
&&a=a_0^{\frac13},\,b=a_0^{-\frac23},\,\,a_0=\frac{2u}{\kappa\sin\psi}~.
\label{III10}
\end{eqnarray}
The difference between $\sigma_{\perp}$ and $\sigma_\parallel$ in
Eqs.~(\ref{III8}), \ref{III9} leads to the asymmetry
\begin{eqnarray}
{\cal A}_\infty=\frac{\sigma_{\perp\,\infty}-\sigma_{\parallel\,\infty}}
                     {\sigma_{\perp\,\infty}+\sigma_{\parallel\,\infty}}~.
\label{III11}
\end{eqnarray}
In the limit of extremely large values of $\zeta\gg2\xi$
(small values of $\kappa\ll1$) and small values of $\zeta\gg2\xi$
(large values of $\kappa\gg1$),
the cross sections take the asymptotic forms
\begin{eqnarray}
&&\zeta\gg2\xi,\,\,\rm or\,\,\,\kappa\ll1\nonumber\\
&&\sigma_{\perp\infty}
=C_\zeta\,\zeta^{-\frac12}{\rm e}^{-\frac{4}{3}\frac{\zeta}{\xi}}
=C_\kappa\,\kappa^{\frac12}{\rm e}^{-\frac{8}{3\kappa}}
,\,\,\,
\sigma_{\parallel\infty}=\frac12\sigma_{\perp\infty},
\nonumber\\
&&
C_\zeta=3\sqrt{\frac\pi\xi}\frac{\alpha^2}{m^2},\qquad
C_\kappa=3\sqrt{\frac\pi2}\frac{\alpha^2}{m^2\xi}~,
\label{III12}\\
&&\zeta\ll2\xi\,\,\rm or\,\,\,\kappa\gg1,
\nonumber\\
&&\sigma_{\perp\infty}
=D_\zeta\,\zeta^\frac13,\
=D_\kappa\,\kappa^{-\frac13},
\,\,\,
\sigma_{\parallel\infty}=\frac23\sigma_{\perp\,\infty},
\nonumber\\
&&
D_\zeta=\frac{3^{\frac{14}{3}}\Gamma^7(\frac32)\,\alpha^2}
     {7\pi^3\xi^{\frac43} m^2},\qquad
D_\kappa=\frac{2^{\frac13}\,3^{\frac{14}{3}}\Gamma^7(\frac32)\,\alpha^2}
     {7\pi^3\xi m^2}~.
\label{III13}
\end{eqnarray}
This leads to the asymptotic expressions for the asymmetry
\begin{eqnarray}
{\cal A}_{\,\zeta\gg2\xi}=\frac13,\qquad\qquad
{\cal A}_{\,\zeta\ll2\xi}=\frac15~.
\label{III14}
\end{eqnarray}

\section{Numerical results, \boldmath$\xi \gg 1$}

\begin{figure}[th]
\includegraphics[width=0.47\columnwidth]{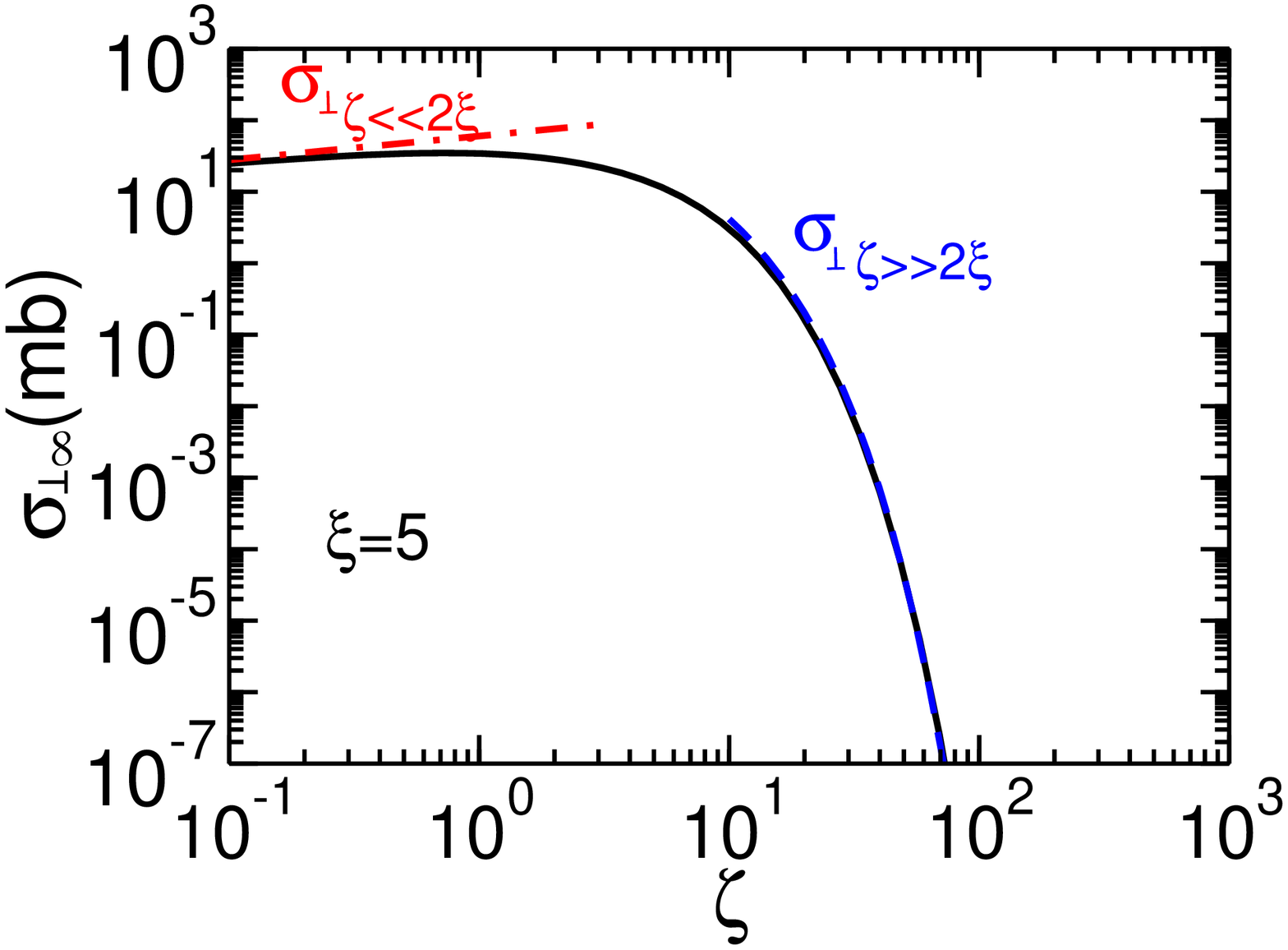} \hfill
\includegraphics[width=0.47\columnwidth]{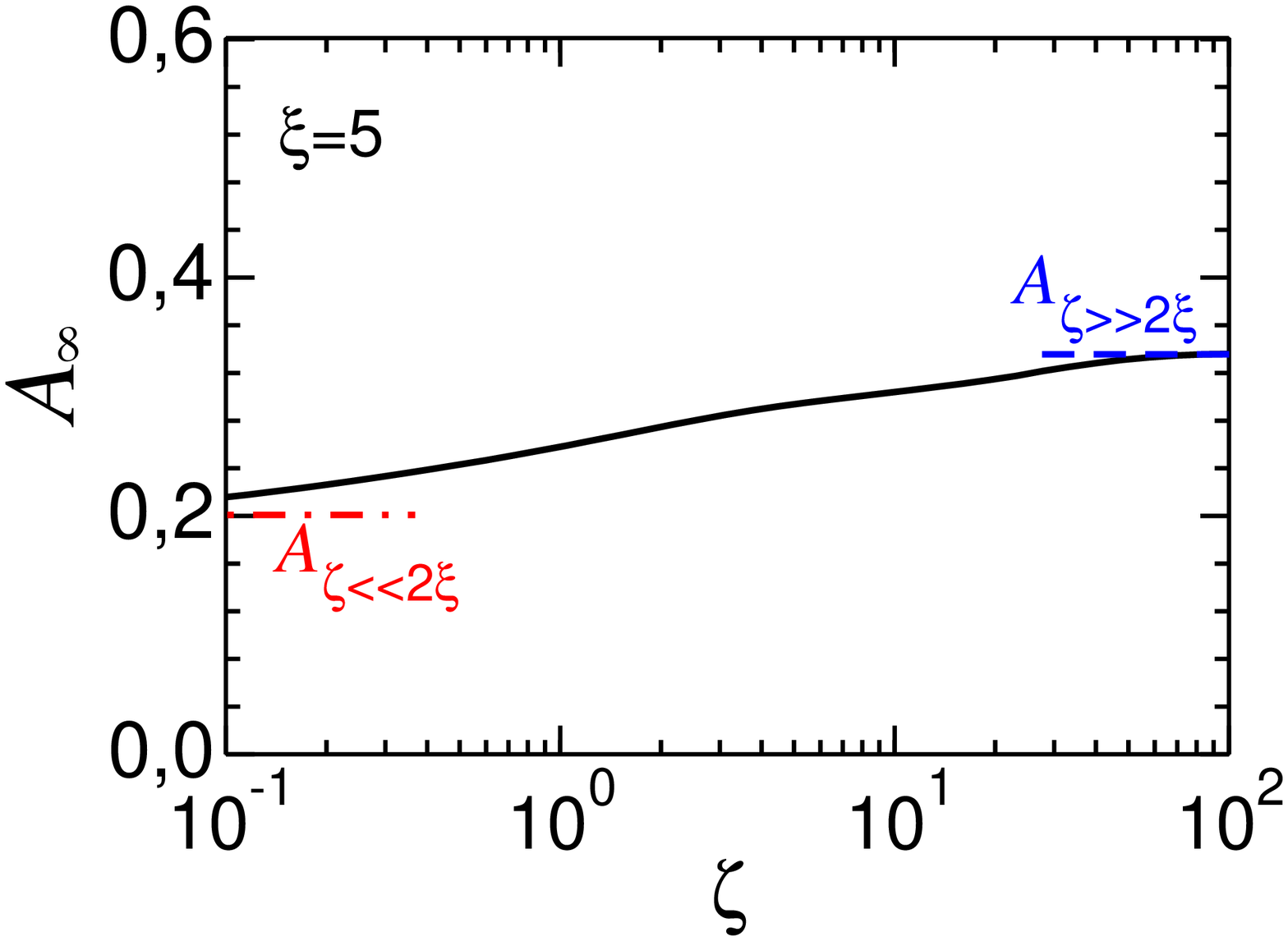}\\
\includegraphics[width=0.47\columnwidth]{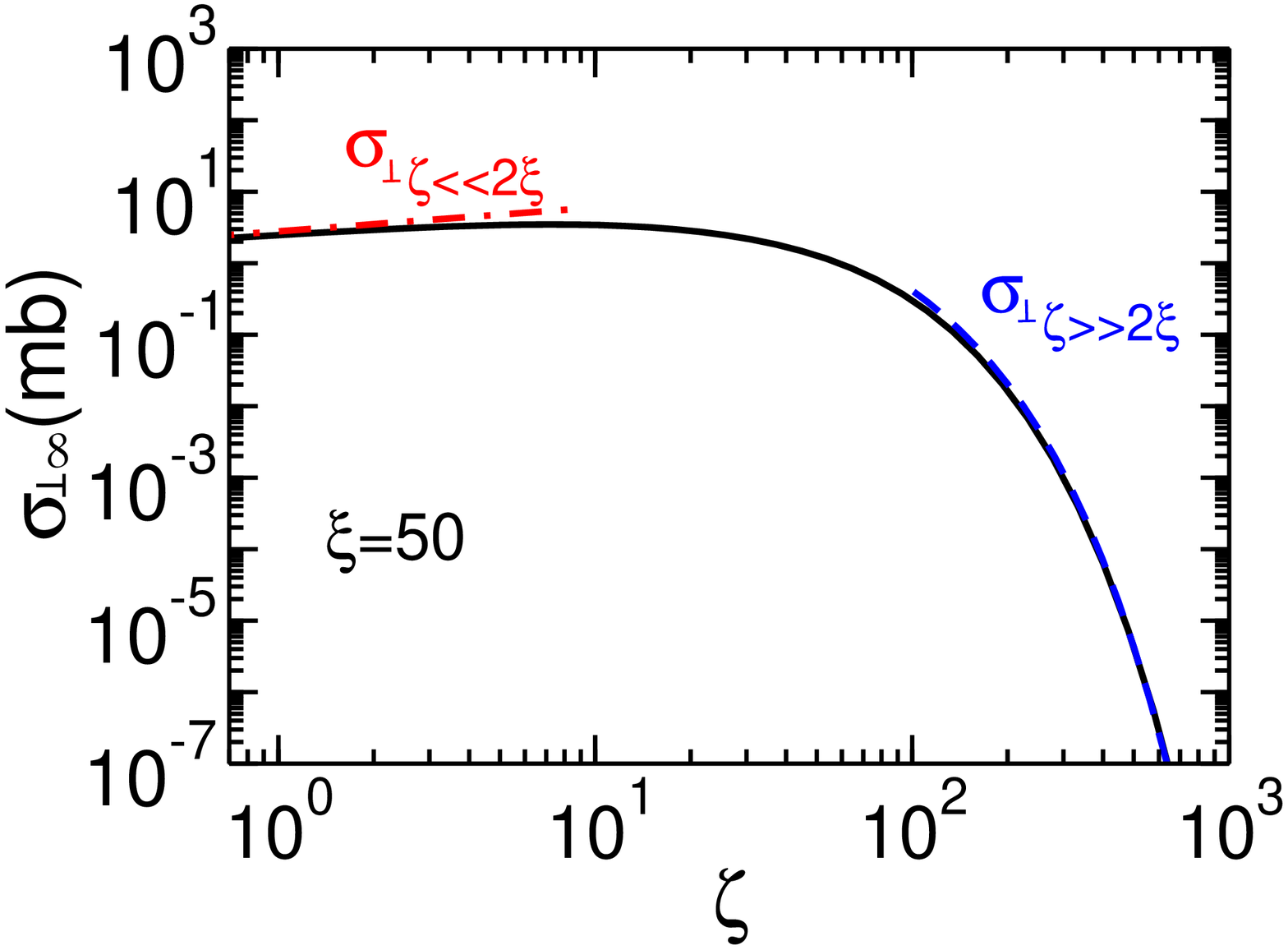} \hfill
\includegraphics[width=0.47\columnwidth]{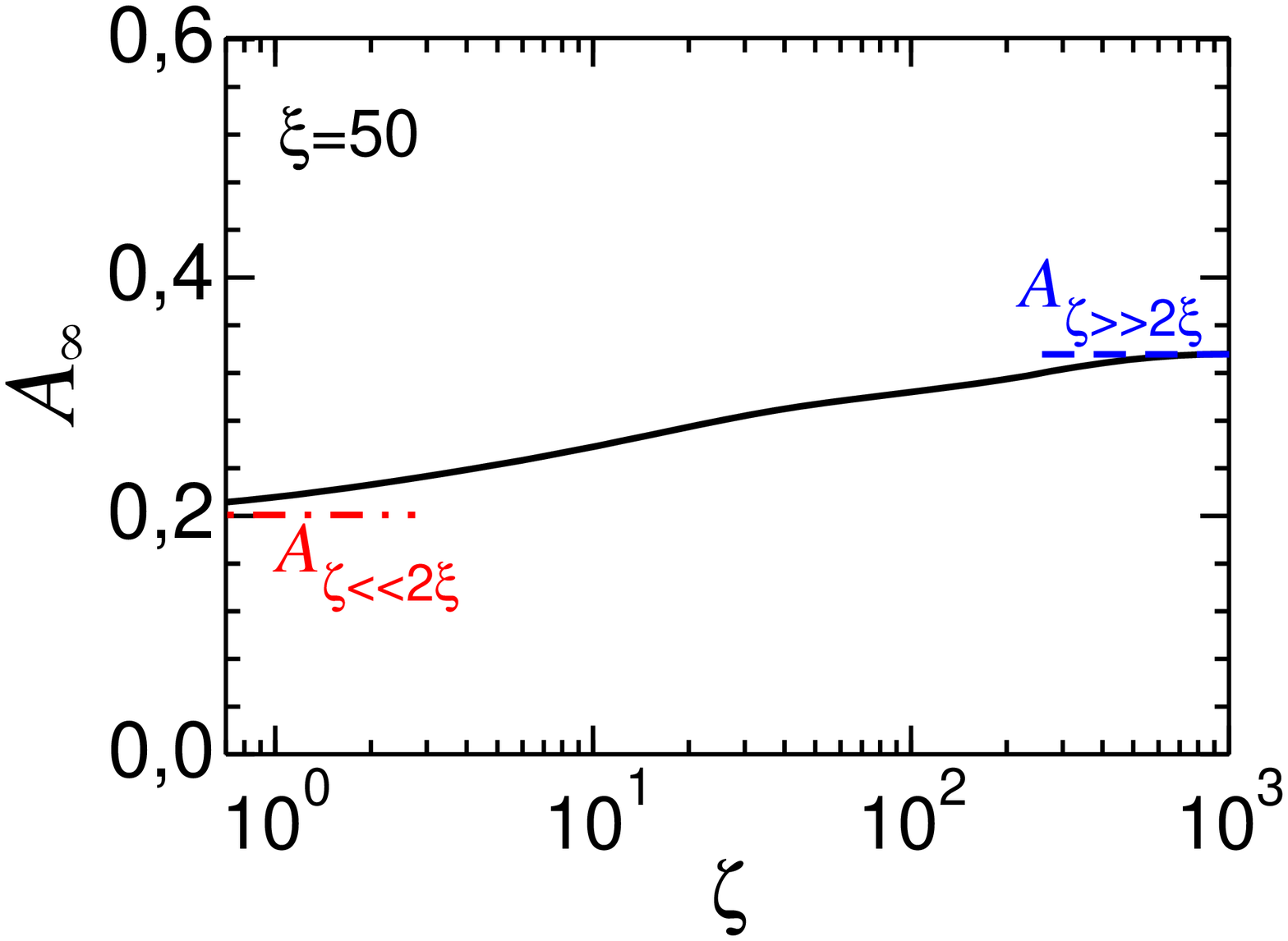}
  \caption{\small{(Color online)
  The left and right panels are for the total
  cross section $\sigma_{\perp\,\infty}$ and asymmetry ${\cal A}_\infty$ as
  a function of $\zeta$. The dashed and dot-dashed curves are for the
  asymptotic expressions Eqs.~(\ref{III12}) and (\ref{III13}), respectively.
  The top and bottom panels are for $\xi=5$ and $50$, respectively.
 \label{Fig:07}}}
 \end{figure}
The total cross sections $\sigma_{\perp\,\infty}$ and the asymmetry ${\cal A}_\infty$
are exhibited in Fig.~\ref{Fig:07} in the left and right panels,
respectively. Results for $\xi=5$ and $50$ are displayed in the
top and bottom panels, respectively.
The dashed and dot-dashed curves are for the asymptotic expressions
Eqs.~(\ref{III12}) and (\ref{III13}), respectively.
At small values of $\zeta$, the cross sections increase slightly from
their asymptotic values and then rapidly decrease with increasing
$\zeta$, being nevertheless finite,
even at very large values of $\zeta\gg1$.

The asymmetry monotonically increases with increasing values of $\zeta$,
being in the range of its asymptotic values
\begin{eqnarray}
0.2\ge A_\infty\ge\frac13 .
\label{III15}
\end{eqnarray}
\begin{figure}[th]
\includegraphics[width=0.47\columnwidth]{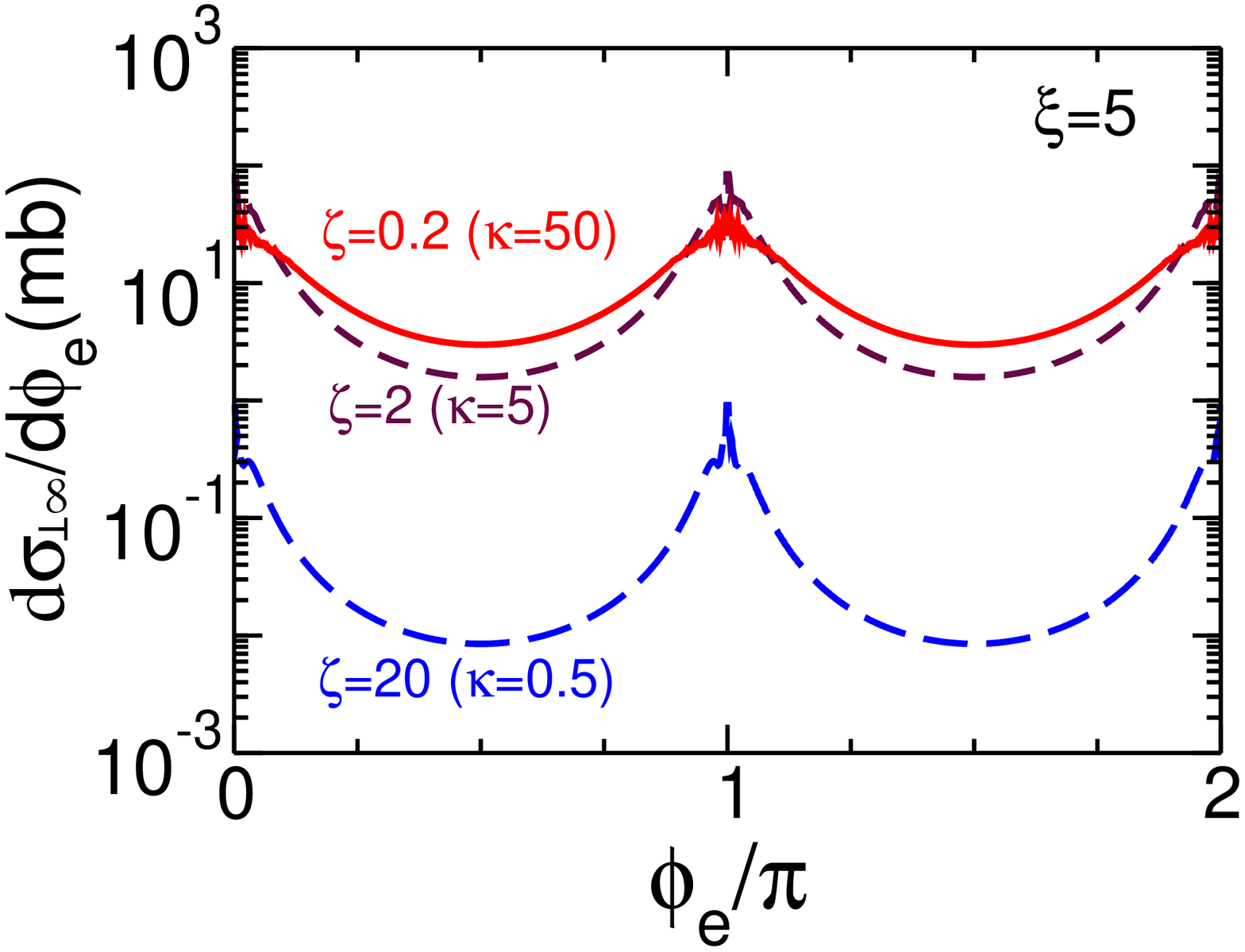} \hfill
\includegraphics[width=0.47\columnwidth]{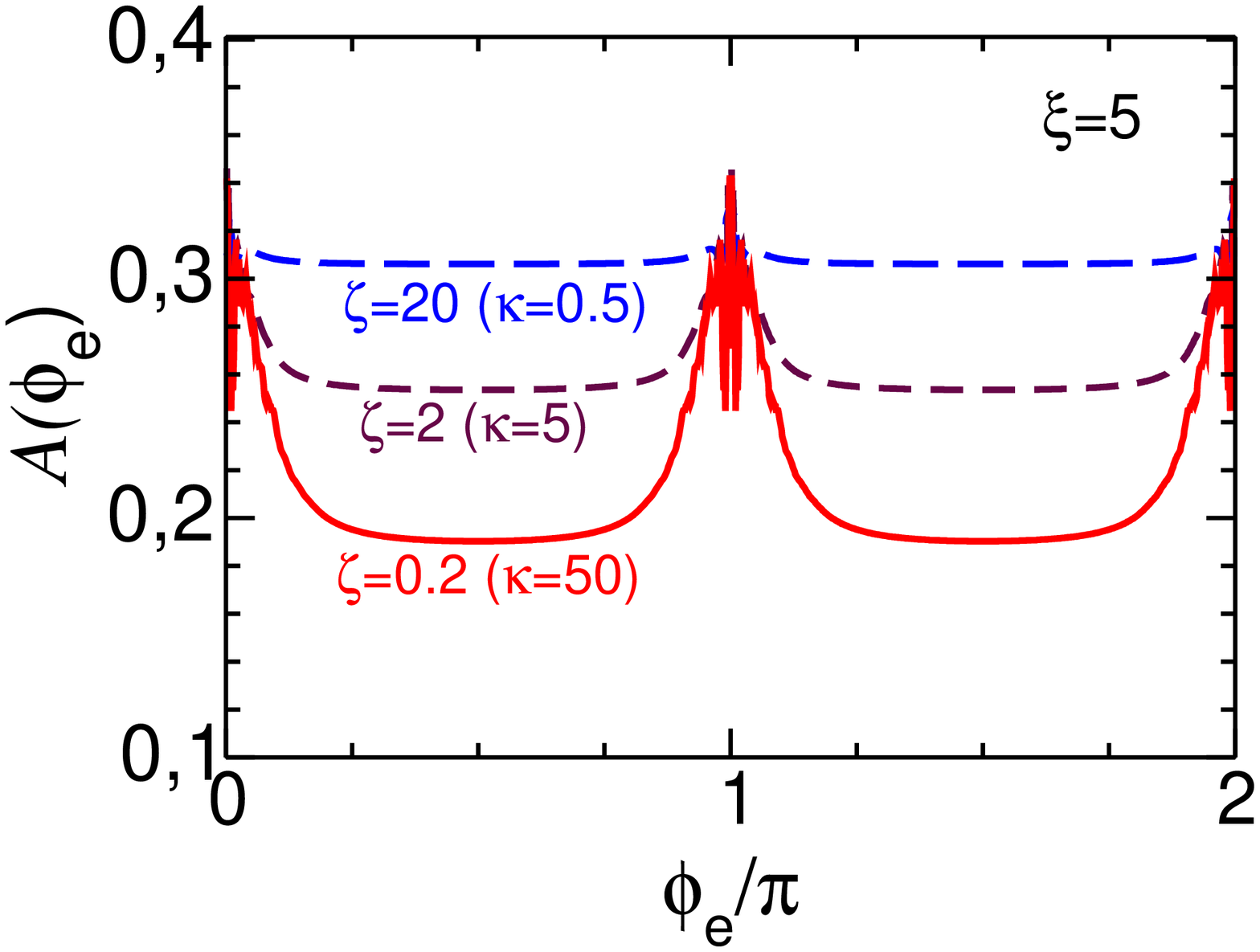}\\
\includegraphics[width=0.47\columnwidth]{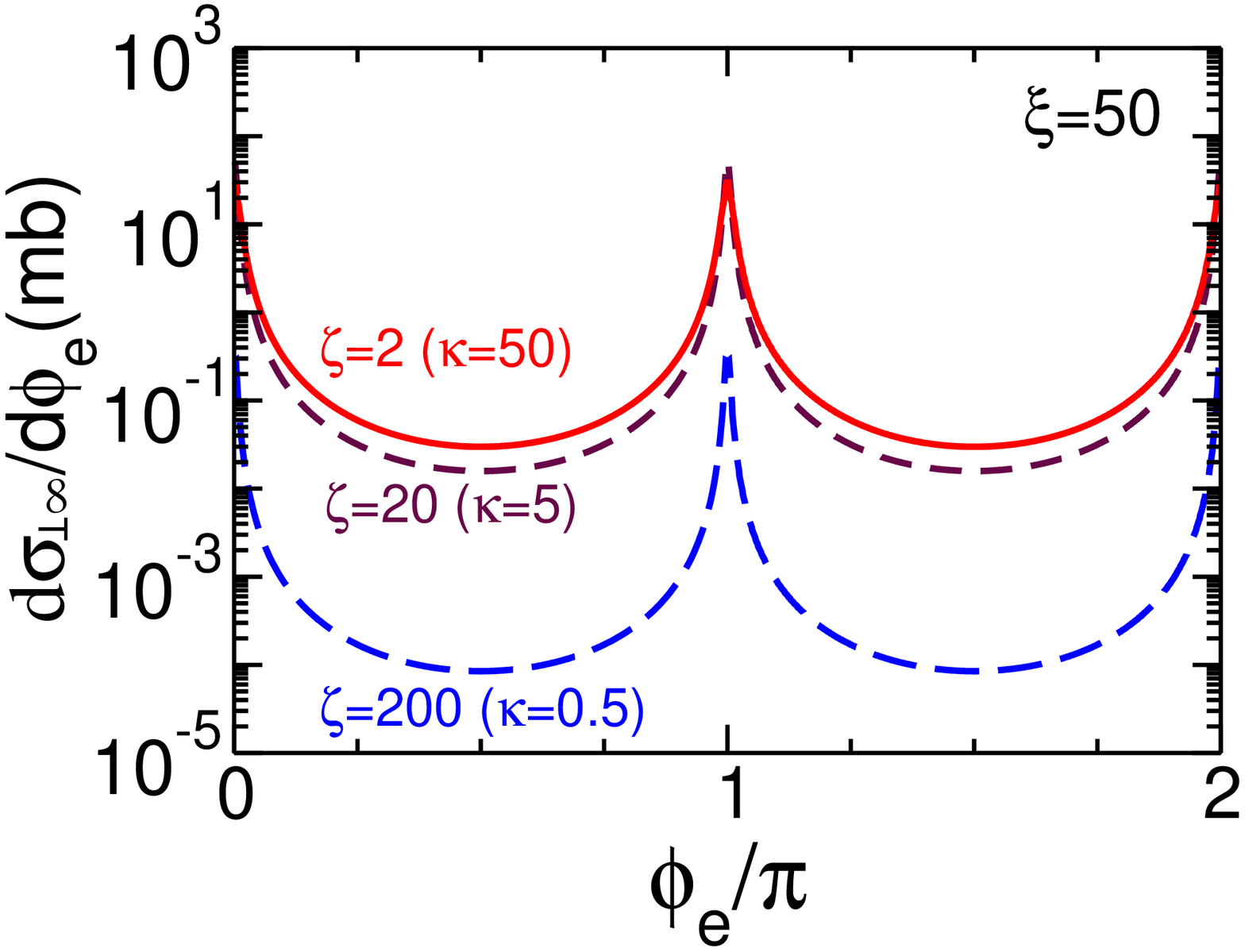} \hfill
\includegraphics[width=0.47\columnwidth]{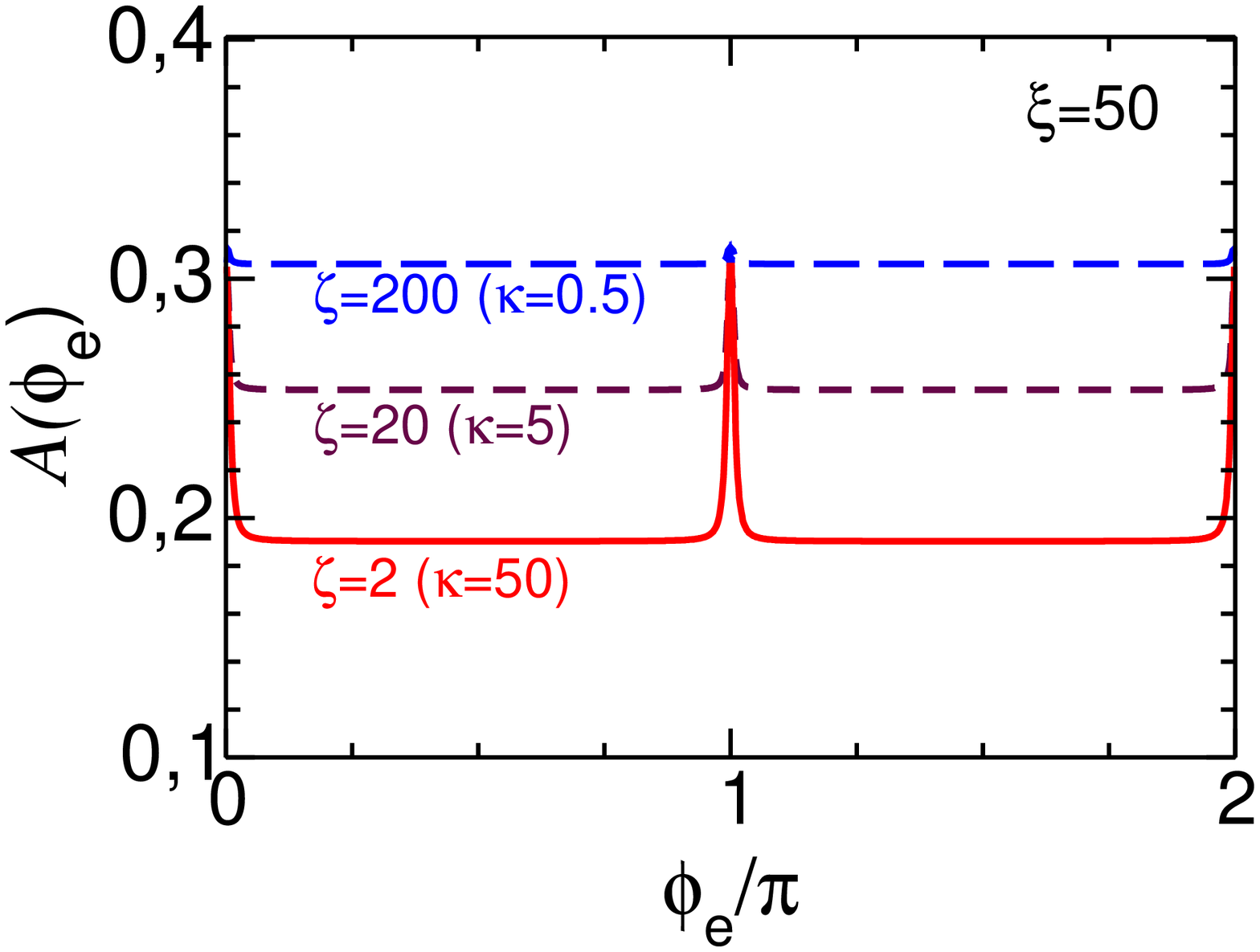}
   \caption{\small{(Color online)
  The left and right panels are for the differential
  cross sections $d\sigma_{\perp\,\infty}/d\phi_e$ and the
  asymmetry ${\cal A}_\infty(\phi_e)$ as a function of the azimuthal angle
  $\phi_e$ for different values of  $\zeta\,(\kappa)$.
  The top and bottom panels are for $\xi=5$ and $50$, respectively.
 \label{Fig:08}}}
 \end{figure}
Using the identity
\begin{eqnarray}
\cos\psi=\frac{\tau}{\xi}\cot\phi~,
\label{III16}
\end{eqnarray}
one can find expressions for the azimuthal angle-differential cross section
\begin{eqnarray}
\hspace{-2mm}\frac{d\sigma_{i\,\infty}}{d\phi}=
\frac{32\alpha^2}{\pi^2\,\xi^2\kappa\,m^2\sin^2\phi}
\int\limits_0^\infty
\frac{dt}{(t^2+1)^\frac32}
\int\limits_{-\infty}^{\infty}
\frac{\vert\tau\vert d\tau}{\sin\psi}w_{i},
\label{III17}
\end{eqnarray}
and the azimuthal angle asymmetry
\begin{eqnarray}
{\cal A}_\infty(\phi)=
\frac{d\sigma_{\perp\,\infty}/d\phi-d\sigma_{\parallel\,\infty}/d\phi}
{d\sigma_{\perp\,\infty}/d\phi+d\sigma_{\parallel\,\infty}/d\phi} .
\label{III18}
\end{eqnarray}
The differential
cross sections $d\sigma_{\perp\,\infty}/d\phi_e$ and
asymmetry ${\cal A}_\infty(\phi_e)$ as a function of the azimuthal angle
$\phi_e$ for different values of $\zeta\,(\kappa)$ are depicted in
Fig.~\ref{Fig:08} in the left and right panels, respectively.
The cross sections exhibit a deep minimum at $\phi_e=\pi/2$ and sharp
maxima at $\phi_e=0,\,\pi$, respectively. The cross sections decrease
with increasing values of $\zeta$.

The asymmetry has sharp peaks at $\phi_e=0,\,\pi$.
The height of the peaks increases with decreasing $\zeta$.
The value of asymmetry in the region
$0<\phi_e<\pi$ is consistent with asymptotic prediction Eq.~(\ref{III14}).

\section{Summary}

In summary we have performed an analysis of the asymmetry
of $\ee$ pair production by the
non-linear BW process for different mutual orientations of the
polarization vectors of a linearly polarized initial probe photon
and the linearly polarized laser pulse for low ($\xi \leq 1$)
and high ($\xi \gg 1$) laser field intensities.
In particular, we examined the asymmetry
caused by the difference of $\spt$ and $\spl$.
We have analyzed the asymmetry for both the total cross sections as a function
of the threshold parameter $\zeta$ and the differential cross sections as a function
of the azimuthal electron angle $\phi_e$ for fixed values of $\zeta$.
Our results can be summarized as follows.

(1) Weak field intensity with $\xi \le 1$:\\
(i) The cross sections $\sigma_{\perp}$
and $\sigma_{\parallel}$ decrease fast with increasing
threshold variable $\zeta$. The cross sections are sensitive
to the pulse duration. Thus,
at relatively large laser pulse duration,
$N\ge 5$,  $\sigma_{\perp, \parallel}$
exhibit a step-like behavior, similar to the prediction
for an infinite pulse. The height of steps is $\propto \xi^{- 2}$.
In case of short and sub-cycle pulses with $N\leq1$,
the cross sections decrease monotonically with increasing $\zeta$.
At $\xi=1$ we found a weak dependence on the pulse duration for $N \geq 1$.\\
(ii) The difference between $\sigma_{\perp}$
and $\sigma_{\parallel}$ generates a specific asymmetry, both
for monochromatic laser beams (IPA) and pulses of finite duration (FPA).
In IPA it has sharp peaks and dips
at integer odd and even values of $\zeta$, respectively.
That is a consequence of properties of the
corresponding basic functions $A_m$. In FPA and for pulses with
$N\geq5$, the asymmetry also exhibits a non-monotonic behavior with
pronounced peaks and dips. Their positions resemble to that of IPA.
At small values of $N$, $N \leq 1$, the asymmetry is a smooth monotonic function
of $\zeta$.\\
(iii) The azimuthal angle dependence of the asymmetries displays smooth
non-monotonic distributions with specific maxima and minima which
are determined by the pulse duration and the threshold parameter $\zeta$.

(2) High laser field intensity with $\xi \gg 1$: \\
(i) The cross sections decrease monotonically with increasing values of $\zeta$
(or decreasing $\kappa$).
At asymptotically small and large $\zeta$ they coincide with
the asymptotic prediction of \cite{Ritus-79}.\\
(ii) The asymmetry increases smoothly with increasing $\zeta$ from $1/5$ to
$1/3$ for $\zeta\ll1$ and $\zeta\gg1$, respectively.\\
(iii) The azimuthal angle distribution exhibits sharp peaks at $\phi_e=0,\,\pi$.
The height of the peaks increases with decreasing values of $\zeta$.

 Our theoretical predictions in a wide region of field intensities
 may be used as a unique and powerful input for the design
 of forthcoming experiments in the near future
 and corresponding high-precision experimental studies of the various
 aspects of multi-photon dynamics in non-linear QED processes.\\

{\it After  completion  of  our  work  we became aware of a paper by D. Seipt
and B. King in [33]“Spin and polarisation dependent LCFA rates for
nonlin-ear Compton and Breit-Wheeler processes”, where further
polarization effects are studied.}

\section*{Acknowledgments}

The authors gratefully acknowledge the collaboration with D. Seipt,
T. Nousch, T. Heinzl,
and useful discussions with A. Ilderton, K. Krajewska,
M. Marklund, C. M\"uller, and R.~Sch\"utzhold.
A.~Ringwald is thanked for explanations w.r.t.\ LUXE. The work is supported
by R. Sauerbrey and T. E. Cowan w.r.t.\ the study of fundamental
QED processes for HIBEF.

\section*{Contributions}

The authors have contributed equally to the publication,
being variously involved in the conceptual outline,
software development and numerical evaluations.


\end{document}